\pdfoutput=1
\RequirePackage{fixltx2e}
\documentclass[11pt,a4paper]{article}
\usepackage[utf8]{inputenc}
\usepackage{jheppub}
\hypersetup{unicode=true,bookmarksopen=true}

\usepackage{mathtools}
\usepackage{lmodern}
\usepackage[T1]{fontenc}
\usepackage{bbm}
\DeclareMathAlphabet{\mathbfi}{OML}{cmm}{b}{it}

\usepackage{subcaption}

\renewcommand{\vec}[1]{{\ifnum9<1#1\mathbf{#1}\else\ifcat\noexpand#1\relax\boldsymbol{#1}\else\mathbfi{#1}\fi\fi}}
\newcommand{\mathe}{\mathrm{e}}
\newcommand{\mathi}{\mathrm{i}}
\let\oldre\Re
\let\oldim\Im
\renewcommand{\Re}{\oldre\mathfrak{e}\,}
\renewcommand{\Im}{\oldim\mathfrak{m}\,}
\newcommand{\total}{\mathop{}\!\mathrm{d}}
\newcommand{\laplace}{\mathop{}\!\bigtriangleup}
\newcommand{\abs}[1]{{\left\lvert{#1}\right\rvert}}

\newcommand{\unitmatrix}{\mathbbm{1}}
\newcommand{\tr}{\operatorname{tr}}
\newcommand{\eqend}[1]{\,#1}
\newcommand{\bigo}[1]{\mathcal{O}\!\left({#1}\right)}

\newcommand{\bra}[1]{\left\langle{#1}\right\vert}
\newcommand{\ket}[1]{\left\vert{#1}\right\rangle}
\newcommand{\brst}{\mathop{}\!\mathsf{s}\hskip 0.05em\relax}
\newcommand{\expect}[1]{\left\langle{#1}\right\rangle}

\newcommand{\hypergeom}[2]{\,{}_{#1}\mathrm{F}_{#2}}

\allowdisplaybreaks

\begin{document}

\title{Quantum gravitational corrections for spinning particles}

\author{Markus B. Fröb}
\affiliation{Department of Mathematics, University of York, Heslington, York, YO10 5DD, United Kingdom}

\emailAdd{mbf503@york.ac.uk}

\abstract{We calculate the quantum corrections to the gauge-invariant gravitational potentials of spinning particles in flat space, induced by loops of both massive and massless matter fields of various types. While the corrections to the Newtonian potential induced by massless conformal matter for spinless particles are well known, and the same corrections due to massless minimally coupled scalars~[\href{http://dx.doi.org/10.1088/0264-9381/27/24/245008}{Class.~Quant.~Grav. \textbf{27} (2010) 245008}], massless non-conformal scalars~[\href{http://dx.doi.org/10.1103/PhysRevD.87.104027}{Phys. Rev.~D \textbf{87} (2013) 104027}] and massive scalars, fermions and vector bosons~[\href{http://dx.doi.org/10.1103/PhysRevD.91.064047}{Phys.~Rev.~D \textbf{91} (2015) 064047}] have been recently derived, spinning particles receive additional corrections which are the subject of the present work. We give both fully analytic results valid for all distances from the particle, and present numerical results as well as asymptotic expansions. At large distances from the particle, the corrections due to massive fields are exponentially suppressed in comparison to the corrections from massless fields, as one would expect. However, a surprising result of our analysis is that close to the particle itself, on distances comparable to the Compton wavelength of the massive fields running in the loops, these corrections can be enhanced with respect to the massless case.}

\keywords{Effective field theories, Models of Quantum Gravity}

\maketitle

\section{Introduction}

While a full theory of quantum gravity is still elusive, and general relativity is non-renormalis\-able as a quantum field theory, certain quantum gravitational predictions can nevertheless be made. Namely, quantising metric fluctuations around a fixed classical background and treating the resulting theory as an effective field theory, one obtains unambiguous predictions whenever the relevant scales of the problem are sufficiently far separated from the fundamental scale where the effective theory breaks down~\cite{donoghue1994b,burgess2004}. Effective field theories have in fact a long history, starting from the Euler-Heisenberg effective Lagrangian for quantum electrodynamics~\cite{heisenbergeuler1936,euler1936}, but their predictive value even in those cases where the underlying fundamental theory is unknown wasn't properly appreciated until the works of Weinberg~\cite{weinberg1979,weinberg1980}. One especially important effect predicted by effective field theories of gravity are quantum corrections to the Newtonian potential, which have been studied by many authors~\cite{radkowski1970,schwinger1968,duff1974,capperduffhalpern1974,capperduff1974,donoghue1994a,donoghue1994b,muzinichvokos1995,hamberliu1995,akhundovbelluccishiekh1997,duffliu2000a,duffliu2000b,kirilinkhriplovich2002,khriplovichkirilin2003,bjerrumbohrdonoghueholstein2003a,bjerrumbohrdonoghueholstein2003b,satzmazzitellialvarez2005,parkwoodard2010,marunovicprokopec2011,marunovicprokopec2012}. The usual way of calculating these corrections is to compute the scattering amplitude for two particles, including loop corrections, and then construct a potential which would produce the same scattering amplitude, i.e., solving the inverse scattering problem. Since scattering amplitudes in flat space are gauge- and reparametrisation-invariant~\cite{kalloshtyutin1972,lam1973}, the resulting potential is as well. At one-loop order and to first order in the mass $M$ of the particle, it reads
\begin{equation}
\label{introduction_vr}
V(r) = - \frac{G_\text{N} M}{r} \left[ 1 + \left( \frac{41}{10 \pi} + \frac{[ 1 + \frac{5}{4} (1-6\xi)^2 ] N_0 + 6 N_{1/2} + 12 N_1}{45 \pi} \right) \frac{\hbar G_\text{N}}{r^2} \right] \eqend{,}
\end{equation}
where $G_\text{N}$ is the Newton constant, the first correction stems from gravitons, $N_s$ is the number of massless spin-$s$ fields running in the loop and $\xi$ determines the non-minimal coupling of the scalar fields to curvature (with conformal coupling being $\xi = 1/6$).

While the inverse scattering technique is well tested and can be easily generalised to higher orders and to the scattering of particles with spin~\cite{holsteinross2008}, the calculation is usually very tedious -- even though modern methods for the computation of scattering amplitudes, such as unitarity or the spinor helicity formalism (see, e.g.,~\cite{bernetal1994,elvanghuang2013,dixon2013}), simplify it, in some cases dramatically. However, there is no obvious generalisation of the inverse scattering technique to curved spaces, where a scattering matrix does not exist in general or, due to horizons, cannot be observed by any single observer~\cite{witten2001,bousso2005}. Fortunately, one can calculate quantum corrections to the Newtonian potential using the same method which is used for the classical calculation and with the same ease, namely by solving the gravitational field equations for a point source~\cite{duff1974,duffliu2000a,duffliu2000b,parkwoodard2010}. These equations naturally cannot come from the classical action, but have to be determined from an effective action which takes into account the vacuum polarisation due to quantum matter. There are various techniques to calculate the effective action, and we will review a particular suitable variant in the next section. Moreover, this approach can also deal with time-dependent sources and backgrounds and provide results for the whole dynamical evolution, while the inverse scattering technique by its very construction is restricted to asymptotic scattering problems. Especially noteworthy in this respect are results for quantum corrections during the inflationary period of the early universe, which are potentially much larger than in flat space due to contributions which grow logarithmically with either time or distance~\cite{wangwoodard2015,parkprokopecwoodard2016,froebverdaguer2016}. Let us finally note that in all cases where the calculation has been done using both methods, they agree completely on the result.

In this article, we take up the question of calculating the quantum corrections to the gravitational potentials of a spinning particle. Using a suitable 3+1-decomposition, one sees that in linearised gravity there are actually four different gauge-invariant potentials (two scalars which one may take to be the flat-space limit of the Bardeen potentials~\cite{bardeen1980}, one transverse vector and a transverse traceless tensor), of which only one scalar potential reduces to the Newtonian potential in the Newtonian limit. For a non-spinning particle, only the scalar potentials are sourced, but even then the quantum corrections are different for both potentials~\cite{duff1974,duffliu2000a,duffliu2000b,parkwoodard2010,marunovicprokopec2011,marunovicprokopec2012,froebverdaguer2016}. While the numerical values of the corrections are practically insignificant, and the Newtonian potential is sufficient to give the correct scattering amplitude, one can in principle construct experiments which are sensitive to the other potentials as well, and which then give a different result from the one obtained by taking only the Newtonian potential into consideration. For spinning particles, also the vector potential (or gravitomagnetic potential) is sourced, which is responsable, e.g., for the Lense-Thirring effect~\cite{lense1918,ciufolinipavlis2004,everittetal2011,iorioetal2011}. For particles with quadrupole or higher moments, one expects that also the tensor potential is sourced, but we do not consider those in the present work. We stress that the calculation presented here is different from one the undertaken in Ref.~\cite{holsteinross2008}: there, the scattering amplitude for two quantum fields of various spins was obtained, while here we study corrections to the potential of a single classic (Lewis-Papapetrou) spinning particle, with arbitrary spin. To connect to the work in Ref.~\cite{holsteinross2008}, one would have to solve the equations of motion for the second (test) particle in the perturbed geometry, which for spinless particles is geodesic motion and for spinning particles has additional spin-spin interactions (see, e.g., Ref.~\cite{lalakpokorskiwess1995}).

The rest of the article is structured as follows: in section~\ref{sec_calculation} we present the calculation of the effective action (including renormalisation) and the corrections to the Newtonian potentials for general matter fields, parametrising the resulting effective action by two non-local kernels. These two kernels, which couple to the linearised Weyl tensor and Ricci scalar, respectively, are then calculated for free massive and massless spin-1 gauge fields, spin-1/2 Dirac fermions and spin-0 scalars in section~\ref{sec_kernels}. For the scalar fields we also include a general coupling to curvature. The results for the quantum-corrected gravitational potentials are then presented in section~\ref{sec_results}, including asymptotic expansions and numerical results (for massive fields). We discuss possible implications and directions for future work in section~\ref{sec_discussion}, and delegate some technical derivations to the appendices.

\section{The calculation}
\label{sec_calculation}

\subsection{Effective action}
\label{sec_calculation_effectiveaction}

The quantum corrections to the gravitational potentials are obtained by solving the field equations coming from an effective action which includes the vacuum polarisation due to quantum matter. This action is the standard one-particle-irreducible effective action obtained by a Legendre transformation. Since we will only consider the vacuum polarisation from matter fields and not gravitons, it is sufficient to expand the gravitational action to second order in perturbations.\footnote{This can be formalised in a large-$N$ expansion, considering $N$ matter fields coupled to gravity and rescaling the Newton constant~\cite{tomboulis1977,hartlehorowitz1981}.} As is well known (or can be easily checked), in this case, and for free (quadratic) theories in general, the effective action is obtained from the classical one by just integrating out the matter fields. Thus, we have
\begin{equation}
\label{calculation_seff_inout}
\exp\left( \mathi S_\text{eff}[h] \right) \equiv \int \exp\left( \mathi S[h,\phi] \right) \mathcal{D} \phi \eqend{,}
\end{equation}
where $h$ denotes the linearised metric perturbation and $\phi$ a general matter field. As usual, the functional integral over the matter fields needs to be regularised, and the proper counterterms included in the total action $S$ such as to make $S_\text{eff}$ finite, and the field equations are obtained by varying $S_\text{eff}$ with respect to the metric perturbation $h_{\mu\nu}$. However, the resulting equations are neither real nor causal, since the path integral in equation~\eqref{calculation_seff_inout} calculates in-out matrix elements instead of true expectation values. The solution is to use the Schwinger-Keldysh or in-in formalism~\cite{schwinger1961,keldysh1964,chousuhaoyu1985,jordan1986}, where one duplicates the set of fields, adding to each usual ``$+$'' field a ``$-$'' partner. For the ``$-$'' fields, time integration is reversed in the action, and equality of both ``$+$'' and ``$-$'' fields is enforced at some final time $T$ which must be larger than any of the times appearing in correlation functions. One can thus view the time integration as running from the initial time, usually taken to be past infinity, to $T$ and back, such that this formalism is also called closed-time-path (CTP) formalism. The ``$+$'' and ``$-$'' labels then just serve to distinguish between the forward and backward part of the contour, and the corresponding path integral calculates $\mathcal{P}$- or path-ordered correlation functions which are the usual time-ordered ones if all fields are ``$+$'', anti-time-ordered if all fields are ``$-$'', and always orders ``$-$'' fields in front of ``$+$'' fields. Thus, in particular,
\begin{equation}
\label{calculation_gmp}
G_{-+}(x,x') \equiv - \mathi \bra{0} \mathcal{P} \phi_-(x) \phi_+(x') \ket{0} = - \mathi \bra{0} \phi(x) \phi(x') \ket{0}
\end{equation}
is the usual (positive frequency) Wightman function, while
\begin{equation}
\label{calculation_gpp}
G_{++}(x,x') \equiv - \mathi \bra{0} \mathcal{P} \phi_+(x) \phi_+(x') \ket{0} = - \mathi \bra{0} \mathcal{T} \phi(x) \phi(x') \ket{0}
\end{equation}
is the Feynman propagator (at tree level). The in-in effective action calculated in this formalism then depends on both ``$+$'' and ``$-$'' metric perturbations and reads
\begin{equation}
\label{calculation_seff_inin}
\exp\left( \mathi S_\text{eff}[h^\pm] \right) \equiv \int \exp\left( \mathi S[h^+,\phi^+] - \mathi S[h^-,\phi^-] \right) \mathcal{D} \phi^\pm \eqend{,}
\end{equation}
where we took the reversal of time integration for the ``$-$'' fields into account by taking the usual action for them with a relative minus sign. The corresponding effective field equations are given by taking a variational derivative with respect to the ``$+$'' fields and setting $h^+ = h^- = h$ afterwards. As we will see (and can be proven in general~\cite{jordan1986}), this gives real and causal evolution equations for the metric perturbation $h$, even though in general they are nonlocal.

Using dimensional regularisation and thus working in $n$ dimensions, we take the action to be the sum of gravitational action, matter action, counterterms and point particle action,
\begin{equation}
S[h,\phi] = S_\text{G}[h] + S_\text{M}[h,\phi] + S_\text{CT}[h] + S_\text{PP}[h] \eqend{,}
\end{equation}
where the gravitational action $S_\text{G}[h]$ is the expansion to second order in metric perturbations off flat space $h_{\mu\nu} \equiv g_{\mu\nu} - \eta_{\mu\nu}$ of the Einstein-Hilbert action
\begin{equation}
S_\text{G} = \frac{1}{\kappa^2} \int R \sqrt{-g} \total^n x
\end{equation}
with $\kappa^2 = 16 \pi G_\text{N}$ with the Newton constant $G_\text{N}$. We parametrise the matter action $S_\text{M}[h,\phi]$ as
\begin{equation}
\label{calculation_matter_action_decomp}
S_\text{M}[h,\phi] = S_\text{M}[\phi] + \frac{1}{2} \int h_{\mu\nu} T^{\mu\nu}[\phi] \total^n x + \iint h_{\mu\nu}(x) h_{\rho\sigma}(y) U^{\mu\nu\rho\sigma}[\phi](x,y) \total^n x \total^n y \eqend{,}
\end{equation}
where $S_\text{M}[\phi]$ is the matter action evaluated in the Minkowski background (which does only contribute an overall unimportant phase factor), $T^{\mu\nu}[\phi]$ is the usual stress tensor, and $U^{\mu\nu\rho\sigma}[\phi]$ is the second variational derivative of the matter action which will (for a local matter action) be proportional to $\delta^n(x-y)$ and its derivatives. The counterterms $S_\text{CT}[h]$ are needed to renormalise the effective action, and are given by the expansion to second order in metric perturbations of
\begin{equation}
\label{calculation_sct}
S_\text{CT} = \delta \frac{\Lambda}{\kappa^2} \int \sqrt{-g} \total^n x + \delta \frac{1}{\kappa^2} \int R \sqrt{-g} \total^n x + \delta \alpha \int C^{\mu\nu\rho\sigma} C_{\mu\nu\rho\sigma} \sqrt{-g} \total^n x + \delta \beta \int R^2 \sqrt{-g} \total^n x \eqend{,}
\end{equation}
where
\begin{equation}
\label{calculation_weyl_def}
C_{\mu\nu\rho\sigma} \equiv R_{\mu\nu\rho\sigma} - \frac{2}{n-2} R_{\mu[\rho} g_{\sigma]\nu} + \frac{2}{n-2} R_{\nu[\rho} g_{\sigma]\mu} + \frac{2}{(n-1)(n-2)} R g_{\mu[\rho} g_{\sigma]\nu}
\end{equation}
is the $n$-dimensional Weyl tensor. Note that because of the Gau{\ss}-Bonnet identity in four dimensions, which in (perturbed) flat space reads
\begin{equation}
\label{calculation_gaussbonnet}
\int \left( R^{\mu\nu\rho\sigma} R_{\mu\nu\rho\sigma} - 4 R^{\mu\nu} R_{\mu\nu} + R^2 \right) \sqrt{-g} \total^n x = 0 \eqend{,}
\end{equation}
we only need two terms quadratic in the curvature, which for convenience we have taken to be the square of the Weyl tensor and the Ricci scalar. Finally, the point particle action is given by
\begin{equation}
S_\text{PP}[h] = \frac{1}{2} \int h_{\mu\nu} T_\text{PP}^{\mu\nu} \total^n x \eqend{,}
\end{equation}
where $T_\text{PP}^{\mu\nu}$ is the point-particle stress tensor whose detailed form we give later. Since we only want to calculate the corrections to the gravitational potentials of the particle, we neglect the backreaction of the particle to the perturbed geometry. Since the backreaction is a higher-order correction, it is sufficient to take the particle action to first order in the perturbation as we have done.

Inserting the action into the definition of the in-in effective action~\eqref{calculation_seff_inin} and expanding the exponentials up to quadratic order in the metric perturbation $h_{\mu\nu}$, we obtain (up to terms which we may ignore since they are independent of $h_{\mu\nu}$)
\begin{equation}
\label{calculation_seff_inin_bare}
\begin{split}
S_\text{eff}[h^\pm] &= S_\text{G}[h^+] - S_\text{G}[h^-] + S_\text{CT}[h^+] - S_\text{CT}[h^-] + S_\text{PP}[h^+] - S_\text{PP}[h^-] \\
&\quad+ \frac{1}{2} \int h^+_{\mu\nu} \expect{ T^{\mu\nu}[\phi^+] }_\phi \total^n x + \iint h^+_{\mu\nu}(x) h^+_{\rho\sigma}(y) \expect{ U^{\mu\nu\rho\sigma}[\phi^+](x,y) }_\phi \total^n x \total^n y \\
&\quad- \frac{1}{2} \int h^-_{\mu\nu} \expect{ T^{\mu\nu}[\phi^-] }_\phi \total^n x - \iint h^-_{\mu\nu}(x) h^-_{\rho\sigma}(y) \expect{ U^{\mu\nu\rho\sigma}[\phi^-](x,y) }_\phi \total^n x \total^n y \\
&\quad+ \frac{\mathi}{8} \iint h^+_{\mu\nu}(x) h^+_{\rho\sigma}(y) \left[ \expect{ T^{\mu\nu}[\phi^+](x) T^{\rho\sigma}[\phi^+](y) }_\phi - \expect{ T^{\mu\nu}[\phi^+](x) }_\phi \expect{ T^{\rho\sigma}[\phi^+](y) }_\phi \right] \total^n x \total^n y \\
&\quad+ \frac{\mathi}{8} \iint h^-_{\mu\nu}(x) h^-_{\rho\sigma}(y) \left[ \expect{ T^{\mu\nu}[\phi^-](x) T^{\rho\sigma}[\phi^-](y) }_\phi - \expect{ T^{\mu\nu}[\phi^-](x) }_\phi \expect{ T^{\rho\sigma}[\phi^-](y) }_\phi \right] \total^n x \total^n y \\
&\quad- \frac{\mathi}{8} \iint h^+_{\mu\nu}(x) h^-_{\rho\sigma}(y) \left[ \expect{ T^{\mu\nu}[\phi^+](x) T^{\rho\sigma}[\phi^-](y) }_\phi - \expect{ T^{\mu\nu}[\phi^+](x) }_\phi \expect{ T^{\rho\sigma}[\phi^-](y) }_\phi \right] \total^n x \total^n y \\
&\quad- \frac{\mathi}{8} \iint h^-_{\mu\nu}(x) h^+_{\rho\sigma}(y) \left[ \expect{ T^{\mu\nu}[\phi^-](x) T^{\rho\sigma}[\phi^+](y) }_\phi - \expect{ T^{\mu\nu}[\phi^-](x) }_\phi \expect{ T^{\rho\sigma}[\phi^+](y) }_\phi \right] \total^n x \total^n y \eqend{,}
\end{split}
\end{equation}
where we defined
\begin{equation}
\expect{ A[\phi] }_\phi \equiv \frac{\int \exp\left( \mathi S_\text{M}[\phi^+] - \mathi S_\text{M}[\phi^-] \right) A[\phi] \mathcal{D} \phi^\pm}{\int \exp\left( \mathi S_\text{M}[\phi^+] - \mathi S_\text{M}[\phi^-] \right) \mathcal{D} \phi^\pm} \eqend{.}
\end{equation}
The divergences that are obtained when taking the expectation values $\expect{ \cdot }_\phi$ must now be absorbed in the counterterms contained in $S_\text{CT}[h^\pm]$. For this, two points are crucial: first, since the counterterm action only gives ``$++$'' and ``$--$'' contributions, that the ``$+-$'' and ``$-+$'' terms in the last two lines are not divergent; and second, that the given counterterms suffice to cancel all divergences, i.e., that the effective theory is renormalisable at this order. The first point is guaranteed by the in-in formalism, essentially because the mixed expectation values involve the Wightman function~\eqref{calculation_gmp} which is not divergent at coincidence, as we will see later on in concrete examples. The second point can be shown nicely using the background field formalism~\cite{dewitt1967,klubergsternzuber1975,arefevafaddeevslavnov1975,abbott1981}: the basic argument is that, since the gauge invariance of the metric perturbations (following from diffeomorphism invariance of the full theory) is unbroken at the quantum level, the counterterms in any regularisation which respects the gauge symmetry, such as dimensional regularisation, must be invariant as well, i.e., scalars constructed out of curvature tensors. Power counting then determines which of those may appear at any given loop order, and at one loop the counterterms shown here are sufficient.

What these arguments do not cover are possible finite terms which remain after subtracting the divergences from the expectation values $\expect{ \cdot }_\phi$. For a general quantum state, these terms must be taken into account, but for the Minkowski vacuum, some of them can be absorbed in the counterterms as well. This fact is non-trivial, but follows from the maximal symmetry of the vacuum state, which leads, e.g., to
\begin{equation}
\expect{ T^{\mu\nu}[\phi] }_\phi = c \eta^{\mu\nu}
\end{equation}
with a constant $c$ which contains both infinite and finite parts. Thus, if the infinite parts can be absorbed into a counterterms, the finite parts can as well (which in this case is a renormalisation of the cosmological constant $\delta \Lambda/\kappa^2$), and similarly for $\expect{ U^{\mu\nu\rho\sigma}[\phi] }_\phi$, which we recall gives rise to a local counterterm since it is proportional to $\delta^n(x-y)$ and its derivatives. In fact, in order to have a renormalised expansion around flat space we must set the renormalised cosmological constant $\Lambda = 0$, which means that it is necessary to absorb all of the finite part in the counterterm $\delta \Lambda/\kappa^2$ as well. Similarly, in order that $G_\text{N}$ (or alternatively $\kappa^2 = 16 \pi G_\text{N}$) corresponds to the renormalised, measured Newton constant, we have to absorb all finite parts in the counterterm $\delta \kappa^{-2}$, such that the coefficient proportional to $R$ in the effective action~\eqref{calculation_seff_inin_bare} is exactly $1/\kappa^2$. In the following, we thus take the quantum state for the matter fields to be the Minkowski vacuum and absorb also the finite parts in the counterterms.

Of course, while the local divergences appearing in the ``$++$'' and ``$--$'' stress tensor two-point functions are canceled by the counterterms proportional to $\delta \alpha$ and $\delta \beta$~\eqref{calculation_sct}, the non-local contributions from these two-point functions (as well as the ``$-+$'' and ``$+-$'' ones) cannot be absorbed, and it is those which give rise to the (in principle) observable corrections to the gravitational potentials. Since the stress tensor is conserved even in the regularised theory (when using dimensional regularisation),
\begin{equation}
\partial_\mu T^{\mu\nu} = 0 \eqend{,}
\end{equation}
and since the Minkowski vacuum is Lorentz invariant, we can write its (regularised) two-point function in the form
\begin{equation}
\label{calculation_tmunu_2pf}
\expect{ T^{\mu\nu}[\phi](x) T^{\rho\sigma}[\phi](y) }_\phi - \expect{ T^{\mu\nu}[\phi](x) }_\phi \expect{ T^{\rho\sigma}[\phi](y) }_\phi = S^{\mu\nu} S^{\rho\sigma} f_1(x-y) + 2 S^{\mu(\rho} S^{\sigma)\nu} f_2(x-y)
\end{equation}
with two scalar functions $f_i$ and the differential operators
\begin{equation}
\label{calculation_smunu_def}
S_{\mu\nu} \equiv \partial_\mu \partial_\nu - \eta_{\mu\nu} \partial^2 \eqend{,}
\end{equation}
which are identically transverse. Note that since we are calculating the connected two-point functions, the result is independent of the known ambiguities in the definition of $T^{\mu\nu}$ which are given by $T^{\mu\nu} \to T^{\mu\nu} + t^{\mu\nu} \unitmatrix$ with a local tensor $t^{\mu\nu}$ constructed out of curvature tensors (see~\cite{hollandswald2005} for a modern proof). Since this ambiguity is proportional to the unit operator $\unitmatrix$, it drops out of the connected two-point function, and for the same reason the trace anomaly has no influence on the result.

Using the expansions from Appendix~\ref{appendix_metric} and integrating by parts, it follows that
\begin{subequations}
\begin{align}
\iint h^{\mu\nu}(x) h^{\rho\sigma}(y) S_{\mu\nu} S_{\rho\sigma} f(x-y) \total^n x \total^n y &= \iint R(x) R(y) f(x-y) \total^n x \total^n y \eqend{,} \\
\iint h^{\mu\nu}(x) h^{\rho\sigma}(y) S_{\mu(\rho} S_{\sigma)\nu} f(x-y) \total^n x \total^n y &= \iint R^{\mu\nu\rho\sigma}(x) R_{\mu\nu\rho\sigma}(y) f(x-y) \total^n x \total^n y \eqend{,}
\end{align}
\end{subequations}
for an arbitrary function $f(x-y)$, where the right-hand sides must be understood to second order in the metric perturbation. Furthermore, also the Gau{\ss}-Bonnet identity in flat space has -- to second order in the metric perturbation -- a non-local counterpart
\begin{equation}
\label{calculation_gaussbonnet_nonlocal}
\iint \left( R^{\mu\nu\rho\sigma}(x) R_{\mu\nu\rho\sigma}(y) - 4 R^{\mu\nu}(x) R_{\mu\nu}(y) + R(x) R(y) \right) f(x-y) \total^n x \total^n y = 0 \eqend{,}
\end{equation}
and we obtain from the definition of the Weyl tensor~\eqref{calculation_weyl_def} that
\begin{equation}
\begin{split}
&\iint C^{\mu\nu\rho\sigma}(x) C_{\mu\nu\rho\sigma}(y) f(x-y) \total^n x \total^n y \\
&= \iint \left( R^{\mu\nu\rho\sigma}(x) R_{\mu\nu\rho\sigma}(y) - \frac{4}{n-2} R^{\mu\nu}(x) R_{\mu\nu}(y) + \frac{2}{(n-1)(n-2)} R(x) R(y) \right) f(x-y) \total^n x \total^n y \eqend{.}
\end{split}
\end{equation}
Taking everything together, it follows that
\begin{equation}
\begin{split}
&\frac{\mathi}{8} \iint h_{\mu\nu}(x) h_{\rho\sigma}(y) \left[ \expect{ T^{\mu\nu}[\phi](x) T^{\rho\sigma}[\phi](y) }_\phi - \expect{ T^{\mu\nu}[\phi](x) }_\phi \expect{ T^{\rho\sigma}[\phi](y) }_\phi \right] \total^n x \total^n y \\
&= \iint C^{\mu\nu\rho\sigma}(x) C_{\mu\nu\rho\sigma}(y) K^\text{bare}_{C^2}(x-y) \total^n x \total^n y + \iint R(x) R(y) K^\text{bare}_{R^2}(x-y) \total^n x \total^n y
\end{split}
\end{equation}
with the two bare, unrenormalised kernels
\begin{subequations}
\label{calculation_kernel_k_def}
\begin{align}
K^\text{bare}_{C^2}(x) &= \frac{\mathi (n-2)}{4 (n-3)} f_2(x) \eqend{,} \\
K^\text{bare}_{R^2}(x) &= \frac{\mathi}{8} f_1(x) + \frac{\mathi}{4 (n-1)} f_2(x) \eqend{.}
\end{align}
\end{subequations}
These two kernels are nothing else but the spin-2 and spin-0 parts of the graviton self-energy, which for free fields was calculated long ago in the time-ordered (the ``$++$'') case~\cite{capperduff1974,capperduffhalpern1974,capper1974} (see also~\cite{martinverdaguer2000} for a scalar field with general mass and curvature coupling). These works were done in momentum space, where the extraction of the differential operators~\eqref{calculation_tmunu_2pf} just corresponds to a reordering of the $p^\mu$, and the spin-2 and spin-0 parts are the coefficients of the two tensor structures one can form out of the $p^\mu$ and the flat metric $\eta^{\mu\nu}$ which are transverse and have the correct symmetries. However, for our purposes it is vastly more useful to have the kernels in position space, and since we need in addition the ``$+-$'' and ``$-+$'' cases, we will thus rederive them for fields of different spins in the next section.

The divergent parts of the bare kernels $K^\text{bare}_{C^2}$ and $K^\text{bare}_{R^2}$ can now be absorbed by the counterterms proportional to $\delta \alpha$ and $\delta \beta$~\eqref{calculation_sct} for the ``$++$'' and ``$--$'' kernels, obtaining renormalised kernels $K^{++}_{C^2/R^2}$ and $K^{--}_{C^2/R^2}$, while the ``$-+$'' and ``$+-$'' kernels are already finite. We can thus take the unregularised limit $n \to 4$, and the full renormalised effective action then reads
\begin{equation}
\label{calculation_seff_inin_ren}
\begin{split}
S_\text{eff}^\text{ren}[h^\pm] &= \frac{1}{\kappa^2} \int R^+ \sqrt{-g^+} \total^4 x - \frac{1}{\kappa^2} \int R^- \sqrt{-g^-} \total^4 x + \frac{1}{2} \int h^+_{\mu\nu} T_\text{PP}^{\mu\nu} \total^4 x - \frac{1}{2} \int h^-_{\mu\nu} T_\text{PP}^{\mu\nu} \total^4 x \\
&\quad+ \iint C^{+\mu\nu\rho\sigma}(x) C^+_{\mu\nu\rho\sigma}(y) \left[ K^{++}_{C^2}(x-y) + \alpha \delta^4(x-y) \right] \total^4 x \total^4 y \\
&\quad- \iint C^{-\mu\nu\rho\sigma}(x) C^+_{\mu\nu\rho\sigma}(y) K^{-+}_{C^2}(x-y) \total^4 x \total^4 y \\
&\quad- \iint C^{+\mu\nu\rho\sigma}(x) C^-_{\mu\nu\rho\sigma}(y) K^{+-}_{C^2}(x-y) \total^4 x \total^4 y \\
&\quad+ \iint C^{-\mu\nu\rho\sigma}(x) C^-_{\mu\nu\rho\sigma}(y) \left[ K^{--}_{C^2}(x-y) - \alpha \delta^4(x-y) \right] \total^4 x \total^4 y \\
&\quad+ \iint R^+(x) R^+(y) \left[ K^{++}_{R^2}(x-y) + \beta \delta^4(x-y) \right] \total^4 x \total^4 y \\
&\quad- \iint R^-(x) R^+(y) K^{-+}_{R^2}(x-y) \total^4 x \total^4 y \\
&\quad- \iint R^+(x) R^-(y) K^{+-}_{R^2}(x-y) \total^4 x \total^4 y \\
&\quad+ \iint R^-(x) R^-(y) \left[ K^{--}_{R^2}(x-y) - \beta \delta^4(x-y) \right] \total^4 x \total^4 y \eqend{,}
\end{split}
\end{equation}
understood to second order in the perturbation $h_{\mu\nu}$.

\subsection{Effective field equations}
\label{sec_calculation_fieldequations}

The effective field equations are now obtained by taking a variational derivative of the renormalised effective action~\eqref{calculation_seff_inin_ren} with respect to $h^+_{\mu\nu}$ and setting $h^+_{\mu\nu} = h^-_{\mu\nu} = h_{\mu\nu}$ afterwards. Using the expansions from Appendix~\ref{appendix_metric}, we obtain
\begin{equation}
E^{\mu\nu} = 0 \eqend{,}
\end{equation}
where
\begin{equation}
\label{calculation_efe_curv}
\begin{split}
E^{\mu\nu} &\equiv R^{\mu\nu} - \frac{1}{2} R g^{\mu\nu} - \frac{\kappa^2}{2} T_\text{PP}^{\mu\nu} - \kappa^2 \int R(y) \left( \nabla^\mu \nabla^\nu - g^{\mu\nu} \nabla^2 \right) \\
&\qquad\quad\times \left[ K^{++}_{R^2}(x-y) + K^{++}_{R^2}(y-x) - K^{-+}_{R^2}(y-x) - K^{+-}_{R^2}(x-y) + 2 \beta \delta^4(x-y) \right] \total^4 y \\
&\quad+ 2 \kappa^2 \int C^{\mu\rho\nu\sigma}(y) \nabla_\rho \nabla_\sigma \\
&\qquad\quad\times \left[ K^{++}_{C^2}(x-y) + K^{++}_{C^2}(y-x) - K^{-+}_{C^2}(y-x) - K^{+-}_{C^2}(x-y) + 2 \alpha \delta^4(x-y) \right] \total^4 y \eqend{,}
\end{split}
\end{equation}
understood to first order in the perturbation $h_{\mu\nu}$. As explained before, since we neglect the backreaction of the particle on the perturbed geometry it is sufficient to expand the point particle action to first order in $h_{\mu\nu}$, such that $T_\text{PP}^{\mu\nu}$ is to be evaluated on the background.

While in Fourier space the different kernels have a vastly different form, in position space they are very similar~\cite{fordwoodard2005,froeb2013}. In general, they are distributions, singular at the origin $x = y$, and -- for the Minkowski vacuum state that we are considering -- Lorentz-invariant. For a suitable choice of renormalisation conditions (i.e., of the finite parts of the counterterms $\delta \alpha$ and $\delta \beta$), they are then the same functions of the invariant distance $(x-y)^2$, but with a different prescription on how to make the resulting distribution well defined:
\begin{equation}
\label{calculation_kernels_func}
K^{AB}_{C^2/R^2}(x-y) = K_{C^2/R^2}\left[ (x-y)_{AB}^2 \right] \eqend{,} \qquad A,B = \pm \eqend{.}
\end{equation}
The different prescriptions are the limits as $\epsilon \to 0$, understood in the distributional sense (i.e., after integrating with a smooth test function), of
\begin{subequations}
\label{calculation_x2_prescriptions}
\begin{align}
(x-y)_{++}^2 &\equiv \left( \vec{x}-\vec{y} \right)^2 - \left( \abs{x^0-y^0} - \mathi \epsilon \right)^2 \eqend{,} \label{calculation_x2_prescription_pp} \\
(x-y)_{-+}^2 &\equiv \left( \vec{x}-\vec{y} \right)^2 - \left( x^0-y^0 - \mathi \epsilon \right)^2 \eqend{,} \label{calculation_x2_prescription_mp} \\
(x-y)_{+-}^2 &\equiv \left( \vec{x}-\vec{y} \right)^2 - \left( x^0-y^0 + \mathi \epsilon \right)^2 \eqend{,} \label{calculation_x2_prescription_pm} \\
(x-y)_{--}^2 &\equiv \left( \vec{x}-\vec{y} \right)^2 - \left( \abs{x^0-y^0} + \mathi \epsilon \right)^2 \label{calculation_x2_prescription_mm} \eqend{.}
\end{align}
\end{subequations}
Especially, we see that
\begin{equation}
K^{++}_{C^2/R^2}(y-x) = K^{++}_{C^2/R^2}(x-y)
\end{equation}
and
\begin{equation}
K^{-+}_{C^2/R^2}(y-x) = K^{+-}_{C^2/R^2}(x-y) \eqend{,}
\end{equation}
which can be used to simplify the effective field equations.

Expanding then the effective field equations~\eqref{calculation_efe_curv} to first order in the perturbation $h_{\mu\nu}$ and integrating by parts, using that the kernels only depend on the difference $x-y$, we obtain finally
\begin{equation}
\label{calculation_efe}
\begin{split}
E^{\mu\nu} &= - \partial^\rho \partial^{(\mu} h^{\nu)}_\rho + \frac{1}{2} \partial^2 h^{\mu\nu} + \frac{1}{2} \partial^\mu \partial^\nu h + \frac{1}{2} \eta^{\mu\nu} S^{\rho\sigma} h_{\rho\sigma} + \frac{\kappa^2}{2} T_\text{PP}^{\mu\nu} \\
&\qquad+ \kappa^2 \int L_{C^2}(x-y) \left( S^{\mu(\rho} S^{\sigma)\nu} - \frac{1}{3} S^{\mu\nu} S^{\rho\sigma} \right) h_{\rho\sigma}(y) \total^4 y \\
&\qquad+ 2 \kappa^2 \int L_{R^2}(x-y) S^{\mu\nu} S^{\rho\sigma} h_{\rho\sigma}(y) \total^4 y \eqend{,}
\end{split}
\end{equation}
where the operators $S_{\mu\nu}$ are defined by equation~\eqref{calculation_smunu_def} and we set
\begin{subequations}
\label{calculation_kernel_l_def}
\begin{align}
L_{C^2}(x-y) &\equiv K^{++}_{C^2}(x-y) - K^{+-}_{C^2}(x-y) + \alpha \delta^4(x-y) \eqend{,} \\
L_{R^2}(x-y) &\equiv K^{++}_{R^2}(x-y) - K^{+-}_{R^2}(x-y) + \beta \delta^4(x-y) \eqend{.}
\end{align}
\end{subequations}

It is well known that linearised gravity is invariant under the gauge symmetry
\begin{equation}
\label{calculation_hmunu_gauge_trafo}
h_{\mu\nu} \to h_{\mu\nu} + 2 \partial_{(\mu} \xi_{\nu)}
\end{equation}
for any vector $\xi_\mu$, and one easily checks that the effective field equations~\eqref{calculation_efe} are invariant under this symmetry. To simplify the equations further, we single out the time direction and perform a decomposition of $h_{\mu\nu}$ into irreducible components under spatial rotations and translations. This decomposition takes the form~\cite{abramobrandenbergermukhanov1997,nakamura2007,froeb2014}
\begin{equation}
\label{calculation_hmunu_inv_gauge}
h_{\mu\nu} = h^\text{inv}_{\mu\nu} + \mathcal{L}_X \eta_{\mu\nu} = h^\text{inv}_{\mu\nu} + 2 \partial_{(\mu} X_{\nu)} \eqend{,}
\end{equation}
where the gauge-invariant part
\begin{equation}
\label{calculation_hmunu_gaugeinvpart}
h^\text{inv}_{\mu\nu} \equiv 2 \delta^0_\mu \delta^0_\nu \Phi_\text{A} + 2 \left( \eta_{\mu\nu} + \delta^0_\mu \delta^0_\nu \right) \Phi_\text{H} + 2 \delta^0_{(\mu} V_{\nu)} + h^\text{TT}_{\mu\nu}
\end{equation}
does not change under infinitesimal coordinate transformations, while the change of $X_\mu$ under the gauge transformation~\eqref{calculation_hmunu_gauge_trafo} is given by the simple one
\begin{equation}
X_\mu \to X_\mu + \xi_\mu \eqend{.}
\end{equation}
The two scalars $\Phi_\text{A}$ and $\Phi_\text{H}$ are the flat-space analogues of the Bardeen potentials~\cite{bardeen1980}, while $V_\mu$ is a spatial transverse vector (i.e., $V_0 = \partial^\mu V_\mu = 0$) and $h^\text{TT}_{\mu\nu}$ a symmetric, spatial transverse and traceless tensor (i.e., $h^\text{TT}_{0\nu} = \partial^\mu h^\text{TT}_{\mu\nu} = 0 = \eta^{\mu\nu} h^\text{TT}_{\mu\nu}$). These four are the gauge-invariant gravitational potentials that we are interested in.

We now insert the above decompositions~\eqref{calculation_hmunu_inv_gauge} and~\eqref{calculation_hmunu_gaugeinvpart} into the effective field equations~\eqref{calculation_efe}. There are four spatial-scalar equations, obtained from $E^{00}$, $\partial_i E^{0i}$, $\delta_{ij} E^{ij}$ and $\partial_i \partial_j E^{ij}$; two spatial-vector equations, obtained from $E^{0i}$ and $\partial_i E^{ij}$ after subtracting the pure-divergence part; and one spatial-tensor equation, obtained from $E^{ij}$ after subtracting divergence and trace parts. To properly subtract those parts, one needs to use that the point particle stress tensor is conserved, which translates to
\begin{equation}
\partial_i T^{0i} = - T^{00\prime} \eqend{,} \quad \partial_i T^{ij} = - T^{0j\prime} \eqend{,}
\end{equation}
where a prime denotes a time derivative, and one needs to assume that the Laplacian has a unique inverse, e.g., with vanishing boundary conditions at spatial infinity. This will be the case for the point particle, and taking suitable linear combinations of the resulting equations we obtain
\begin{subequations}
\label{calculation_efe_svt_full}
\begin{align}
\begin{split}
\laplace \Phi_\text{A} &= - \frac{\kappa^2}{4} T^\text{(S)} + \frac{2}{3} \kappa^2 \left( \laplace - 3 \partial^2 \right) \int L_{C^2}(x-y) \laplace \left( \Phi_\text{A} + \Phi_\text{H} \right)(y) \total^4 y \\
&\qquad+ 2 \kappa^2 \laplace \int L_{R^2}(x-y) \left( \laplace \Phi_\text{A} - 2 \laplace \Phi_\text{H} + 3 \Phi_\text{H}'' \right)(y) \total^4 y \eqend{,}
\end{split} \\
\begin{split}
\laplace \Phi_\text{H} &= - \frac{\kappa^2}{4} T_\text{PP}^{00} - \frac{2}{3} \kappa^2 \laplace \int L_{C^2}(x-y) \left( \laplace \Phi_\text{A} + \laplace \Phi_\text{H} \right)(y) \total^4 y \\
&\qquad- 2 \kappa^2 \laplace \int L_{R^2}(x-y) \left( \laplace \Phi_\text{A} - 2 \laplace \Phi_\text{H} + 3 \Phi_\text{H}'' \right)(y) \total^4 y \eqend{,}
\end{split} \\
\laplace V_i &= - \kappa^2 T^\text{(V)}_i - 2 \kappa^2 \partial^2 \int L_{C^2}(x-y) \laplace V_i(y) \total^4 y \eqend{,} \\
\partial^2 h^\text{TT}_{ij} &= - \kappa^2 T^\text{(TT)}_{ij} - 2 \kappa^2 \partial^2 \int L_{C^2}(x-y) \partial^2 h^\text{TT}_{ij}(y) \total^4 y \eqend{,}
\end{align}
\end{subequations}
where we defined
\begin{subequations}
\label{calculation_stress_tensor_combi}
\begin{align}
T &\equiv \delta_{ij} T_\text{PP}^{ij} - T_\text{PP}^{00} \eqend{,} \\
T^\text{(S)} &\equiv T_\text{PP}^{00} + \delta_{ij} T_\text{PP}^{ij} + 3 \frac{\partial^k}{\laplace} T_\text{PP}^{0k\prime} \eqend{,} \\
T^\text{(V)}_i &\equiv - \left( \delta_{ij} - \frac{\partial_i \partial_j}{\laplace} \right) T_\text{PP}^{0j} \eqend{,} \\
T^\text{(TT)}_{ij} &\equiv \left( \delta_{ik} \delta_{jl} - \frac{1}{2} \delta_{ij} \delta_{kl} + \frac{1}{2} \frac{\partial_i \partial_j}{\laplace} \delta_{kl} \right) T_\text{PP}^{kl} + \left( 2 \frac{\partial_{(i} \delta_{j)k}}{\laplace} - \frac{1}{2} \delta_{ij} \frac{\partial_k}{\laplace} - \frac{1}{2} \frac{\partial_i \partial_j \partial_k}{\laplace^2} \right) T_\text{PP}^{0k\prime} \eqend{.}
\end{align}
\end{subequations}
Of the four spatial-scalar equations, only two are independent, while the other ones can be obtained from the ones shown by taking time-derivatives. Similarly, only one of the two spatial-vector equations is independent, and shown above. Note that these effective field equations are coupled integro-differential equations, and that they are real and causal due to the support properties of the integrand. We have to distinguish three cases: a) $y$ is in the forward lightcone of $x$, b) $y$ and $x$ and spacelike separated, and c) $y$ is in the backward lightcone of $y$. In case a), we have $y^0 > x^0$ and therefore $(x-y)_{++}^2 = \left( \vec{x}-\vec{y} \right)^2 - \left( -x^0+y^0 - \mathi \epsilon \right)^2 = (x-y)_{+-}^2$ [see equation~\eqref{calculation_x2_prescriptions}], while in case b) we can perform the limit $\epsilon \to 0$ straightforwardly since $\left( \vec{x}-\vec{y} \right)^2 > \left( x^0-y^0 \right)^2$ for spacelike separations, and then also $(x-y)_{++}^2 = (x-y)_{+-}^2$. In both cases, we thus see that $L_{C^2/R^2}(x-y) = 0$. In case c), we have $y^0 < x^0$, which leads to $(x-y)_{++}^2 = \left[ (x-y)_{+-}^2 \right]^*$, and thus $L_{C^2/R^2}(x-y)$ does not vanish, but since the kernels $K_{C^2/R^2}$ have an explicit factor of $\mathi$~\eqref{calculation_kernel_k_def} the difference appearing in the kernels $L_{C^2/R^2}(x-y)$ is real. It is thus explicitly seen how the in-in formalism guarantees real and causal field equations~\cite{jordan1986}.

We can now distinguish two contributions to the gravitational potentials: the first one is entirely classical and is obtained from the full equations~\eqref{calculation_efe_svt_full} taking only the classical stress tensor of the point particle into account, while the second one represents the quantum corrections in which we are interested. As can be seen from the explicit form of the effective field equations~\eqref{calculation_efe_svt_full}, this second contribution is suppressed by an explicit factor of $\kappa^2$, and we thus decompose
\begin{equation}
\label{calculation_efe_classical_quantum_decomp}
\Phi_\text{A} = \Phi_\text{A}^\text{cl} + \kappa^2 \Phi_\text{A}^\text{qu} \eqend{,}
\end{equation}
and analogously for the other gravitational potentials. For the classical contribution, we therefore obtain the equations
\begin{subequations}
\label{calculation_efe_svt_classical}
\begin{align}
\laplace \Phi_\text{A}^\text{cl} &= - \frac{\kappa^2}{4} T^\text{(S)} \eqend{,} \\
\laplace \Phi_\text{H}^\text{cl} &= - \frac{\kappa^2}{4} T_\text{PP}^{00} \eqend{,} \\
\laplace V_i^\text{cl} &= - \kappa^2 T^\text{(V)}_i \eqend{,} \\
\partial^2 h^\text{TT,cl}_{ij} &= - \kappa^2 T^\text{(TT)}_{ij} \eqend{,}
\end{align}
\end{subequations}
which can be solved once the point-particle stress tensor has been specified, which we will do in subsection~\ref{sec_calculation_pointparticle}. It can also be nicely seen that the spatial-scalar and spatial-vector equations are constraint equations, such that the two scalars and the vector are fully determined once the stress tensor has been given, while the tensor contains the dynamical degrees of freedom (besides being sourced by the tensor part of the stress tensor). The quantum contribution is sourced by the classical potentials, and we obtain from the full equations~\eqref{calculation_efe_svt_full} that
\begin{subequations}
\label{calculation_efe_svt_quantum}
\begin{align}
\begin{split}
\laplace \Phi_\text{A}^\text{qu} &= \frac{2}{3} \left( \laplace - 3 \partial^2 \right) \int L_{C^2}(x-y) \laplace \left( \Phi_\text{A}^\text{cl} + \Phi_\text{H}^\text{cl} \right)(y) \total^4 y \\
&\qquad+ 2 \laplace \int L_{R^2}(x-y) \left( \laplace \Phi_\text{A}^\text{cl} - 2 \laplace \Phi_\text{H}^\text{cl} + 3 \Phi_\text{H}^{\text{cl}\prime\prime} \right)(y) \total^4 y \eqend{,}
\end{split} \\
\begin{split}
\laplace \Phi_\text{H}^\text{qu} &= - \frac{2}{3} \laplace \int L_{C^2}(x-y) \left( \laplace \Phi_\text{A}^\text{cl} + \laplace \Phi_\text{H}^\text{cl} \right)(y) \total^4 y \\
&\qquad- 2 \laplace \int L_{R^2}(x-y) \left( \laplace \Phi_\text{A}^\text{cl} - 2 \laplace \Phi_\text{H}^\text{cl} + 3 \Phi_\text{H}^{\text{cl}\prime\prime} \right)(y) \total^4 y \eqend{,}
\end{split} \\
\laplace V_i^\text{qu} &= - 2 \partial^2 \int L_{C^2}(x-y) \laplace V_i^\text{cl}(y) \total^4 y \eqend{,} \\
\partial^2 h^\text{TT,qu}_{ij} &= - 2 \partial^2 \int L_{C^2}(x-y) \partial^2 h^\text{TT,cl}_{ij}(y) \total^4 y \eqend{.}
\end{align}
\end{subequations}

\subsection{Spinning point particle}
\label{sec_calculation_pointparticle}

In the classical formulation of spinning particles within general relativity~\cite{mathisson1937,papapetrou1951,trautman1958,tulczyjew1959,tulczyjew1962,taub1964,dixon1964}, spin is described by an antisymmetric spin tensor $S^{\mu\nu}(\tau)$ in addition to the four-velocity
\begin{equation}
u^\mu(\tau) \equiv \frac{\total z^\mu(\tau)}{\total \tau}
\end{equation}
with $z^\mu(\tau)$ being the position of the particle at proper time $\tau$, and the linear momentum $p^\mu(\tau)$. In absence of spin, we have $p^\mu = M u^\mu$ where $M$ is the mass of the particle, but this does not hold in general if the spin tensor does not vanish. The stress tensor takes then the form (see Refs.~\cite{ohashi2003,steinhoff2010,blanchet2011} for a review)
\begin{equation}
\label{spinning_stress_tensor}
T_\text{PP}^{\mu\nu}(x) = \int \delta(x-z(\tau)) p^{(\mu}(\tau) u^{\nu)}(\tau) \total \tau - \nabla_\alpha \int \delta(x-z(\tau)) S^{\alpha(\mu}(\tau) u^{\nu)}(\tau) \total \tau
\end{equation}
with the covariant $\delta$ distribution
\begin{equation}
\delta(x-y) \equiv \frac{\delta^n(x-y)}{\sqrt{-g(x)}} \eqend{.}
\end{equation}
From its covariant conservation, using that $\total/\total \tau = u^\mu \nabla_\mu$, we find the equation of motion for the particle (the Mathisson-Papapetrou equation), which reads
\begin{equation}
\label{spinning_pp_eom}
\frac{\total p_\alpha}{\total \tau} = - \frac{1}{2} R_{\alpha\beta\mu\nu} u^\beta S^{\mu\nu} \eqend{,}
\end{equation}
and the spin precession equation
\begin{equation}
\label{spinning_pp_spin}
\frac{\total S^{\mu\nu}}{\total \tau} = p^\mu u^\nu - p^\nu u^\mu \eqend{.}
\end{equation}
Given initial conditions, the solution of these equations is only unique if we specify an additional constraint equation for the spin tensor. The ones studied in the literature are the Frenkel-Pirani condition~\cite{frenkel1926,pirani1956}
\begin{equation}
\label{spinning_spin_condition_fp}
S^{\mu\nu} u_\mu = 0
\end{equation}
and the Tulczyjew condition~\cite{tulczyjew1959,tulczyjew1962}
\begin{equation}
\label{spinning_spin_condition_tu}
S^{\mu\nu} p_\mu = 0 \eqend{.}
\end{equation}
Note that for either of these conditions, the spin tensor is conserved in magnitude, as follows from
\begin{equation}
\label{spinning_spin_tensor_conserved}
\frac{\total \left( S^{\mu\nu} S_{\mu\nu} \right)}{\total \tau} = 4 S^{\mu\nu} p_\mu u_\nu = 0 \eqend{.}
\end{equation}

For a background Minkowski spacetime, the Riemann tensor vanishes, and thus the equation of motion reduces to
\begin{equation}
\label{spinning_pp_eom_mink}
\frac{\total p^\mu}{\total \tau} = 0 \eqend{.}
\end{equation}
We are interested in a particle at rest at the origin, such that
\begin{equation}
z^\mu(\tau) = \tau \delta^\mu_0
\end{equation}
and
\begin{equation}
u^\mu(\tau) = \delta^\mu_0 \eqend{,}
\end{equation}
which has the correct normalisation
\begin{equation}
u^\mu u_\mu = -1 \eqend{.}
\end{equation}
Taking then $p^\mu = M u^\mu$ with constant $M$ as in the spinless case, the equations of motion~\eqref{spinning_pp_eom_mink} and~\eqref{spinning_pp_spin} are satisfied for a constant spin tensor $S^{\mu\nu}$. Moreover, both the Frenkel-Pirani~\eqref{spinning_spin_condition_fp} and Tulczyjew conditions~\eqref{spinning_spin_condition_tu} are satisfied. Since $S^{\mu\nu}$ is antisymmetric, we can then alternatively fully characterise the spin of the particle by the spin vector
\begin{equation}
S_\mu \equiv \frac{1}{2} \epsilon_{\mu\nu\rho\sigma} u^\nu S^{\rho\sigma} \eqend{,}
\end{equation}
which is also seen to be constant and purely spatial, i.e., $S_\mu u^\mu = 0$.

For this particle, the components of the stress tensor~\eqref{spinning_stress_tensor} are easily calculated to be
\begin{subequations}
\label{spinning_stress_tensor_components}
\begin{align}
T_\text{PP}^{00} &= M \delta^3(\vec{x}) \eqend{,} \\
T_\text{PP}^{0i} &= - \frac{1}{2} \epsilon^{ijk} S_j \partial_k \delta^3(\vec{x}) \eqend{,} \\
T_\text{PP}^{ij} &= 0 \eqend{,}
\end{align}
\end{subequations}
and since the stress tensor is time-independent, for the combinations~\eqref{calculation_stress_tensor_combi} we obtain
\begin{subequations}
\label{spinning_stress_tensor_combi}
\begin{align}
T^\text{(S)} &= - T = M \delta^3(\vec{x}) \eqend{,} \\
T^\text{(V)}_i &= \frac{1}{2} \epsilon_{ijk} S^j \partial^k \delta^3(\vec{x}) \eqend{,} \\
T^\text{(TT)}_{ij} &= 0 \eqend{.}
\end{align}
\end{subequations}
We note at this point that it is also possible to introduce a non-minimal spin-gravity coupling~\cite{deriglazovramirez2015a,deriglazovramirez2015b}. Similarly to the case of a non-minimally coupled scalar field, this additional coupling does not change the equations of motion for the particle in the present case (geodesic motion in flat space), but gives rise to a modified stress-energy tensor. However, the corrections are of quadratic order in the spin tensor, and working to first order in spin we can neglect them.

The classical field equations for the gravitational potentials~\eqref{calculation_efe_svt_classical} then reduce to
\begin{subequations}
\label{spinning_efe_svt_classical}
\begin{align}
\laplace \Phi_\text{A}^\text{cl} &= \laplace \Phi_\text{H}^\text{cl} = - \frac{\kappa^2}{4} M \delta^3(\vec{x}) \eqend{,} \\
\laplace V_i^\text{cl} &= - \frac{\kappa^2}{2} \epsilon_{ijk} S^j \partial^k \delta^3(\vec{x}) \eqend{,} \\
\partial^2 h^\text{TT,cl}_{ij} &= 0 \eqend{,}
\end{align}
\end{subequations}
and using that
\begin{equation}
\label{spinning_delta_laplace}
\delta^3(\vec{x}) = - \frac{1}{4\pi} \laplace \frac{1}{r}
\end{equation}
with $r \equiv \abs{\vec{x}}$, we obtain the solutions
\begin{subequations}
\label{spinning_efe_sol_classical}
\begin{align}
\Phi_\text{A}^\text{cl} &= \Phi_\text{H}^\text{cl} = \frac{\kappa^2 M}{16 \pi r} \eqend{,} \\
V_i^\text{cl} &= \frac{\kappa^2}{8\pi} \epsilon_{ijk} S^j \partial^k \frac{1}{r} = - \frac{\kappa^2 (\vec{S} \times \vec{r})_i}{8 \pi r^3} \eqend{,} \\
h^\text{TT,cl}_{ij} &= 0 \eqend{.}
\end{align}
\end{subequations}
Using that $\kappa^2 = 16 \pi G_\text{N}$ and taking $X_\mu = 0$ in the decomposition~\eqref{calculation_hmunu_inv_gauge}, we obtain the linearised metric perturbation in the form
\begin{equation}
\begin{split}
h_{\mu\nu} \total x^\mu \total x^\nu &= 2 \frac{G_\text{N} M}{r} \total t^2 + 2 \frac{G_\text{N} M}{r} \total \vec{x}^2 - 4 \frac{G_\text{N} (\vec{S} \times \vec{r})_i}{r^3} \total t \total x^i  \\
&= 2 \frac{G_\text{N} M}{r} \total t^2 + 2 \frac{G_\text{N} M}{r} \left( \total r^2 + r^2 \total \theta^2 + r^2 \sin^2 \theta \total \phi^2 \right) - 4 \frac{\abs{\vec{S}}}{r} \sin^2 \theta \total t \total \phi \eqend{,}
\end{split}
\end{equation}
where the second equality is obtained by switching to the usual spherical coordinates where we have, assuming that $\vec{S}$ is oriented in the $z$-direction, 
\begin{equation}
\frac{(\vec{S} \times \vec{r})_i}{r^3} \total x^i = \frac{\abs{\vec{S}}}{r} \sin^2 \theta \total \phi \eqend{.}
\end{equation}
This is exactly the far-field form of the Kerr metric~\cite{kramerstephani} if we identify the rotation parameter $a$ with
\begin{equation}
\label{spinning_efe_param_a}
a = \frac{\abs{\vec{S}}}{G_\text{N} M}
\end{equation}
and thus the angular momentum $J$ of the Kerr metric with $J = \abs{\vec{S}}$. If we would have taken into account the backreaction of the particle on the geometry, or kept terms of higher order in spin (in the case of a non-minimal spin-gravity coupling), this classical result would obtain corrections of second or higher order in $M$ and $a$. It would be interesting (but beyond the scope of this work) to see if these corrections coincide with a higher-order expansion of the classical Kerr metric.

Since the solutions for the classical gravitational potentials~\eqref{spinning_efe_sol_classical} are time-independent, the sources on the right-hand side of the equations for the quantum contributions~\eqref{calculation_efe_svt_quantum} are also time-independent after the change of integration variable $y \to x-y$. All time derivatives acting on them thus vanish, and after removing an overall Laplacian we obtain
\begin{subequations}
\label{spinning_efe_svt_quantum}
\begin{align}
\begin{split}
\Phi_\text{A}^\text{qu} &= - \frac{4}{3} \int L_{C^2}(y) \laplace \left( \Phi_\text{A}^\text{cl} + \Phi_\text{H}^\text{cl} \right)(x-y) \total^4 y \\
&\qquad+ 2 \int L_{R^2}(y) \laplace \left( \Phi_\text{A}^\text{cl} - 2 \Phi_\text{H}^\text{cl} \right)(x-y) \total^4 y \eqend{,}
\end{split} \\
\begin{split}
\Phi_\text{H}^\text{qu} &= - \frac{2}{3} \int L_{C^2}(y) \laplace \left( \Phi_\text{A}^\text{cl} + \Phi_\text{H}^\text{cl} \right)(x-y) \total^4 y \\
&\qquad- 2 \int L_{R^2}(y) \laplace \left( \Phi_\text{A}^\text{cl} - 2 \Phi_\text{H}^\text{cl} \right)(x-y) \total^4 y \eqend{,}
\end{split} \\
V_i^\text{qu} &= - 2 \int L_{C^2}(y) \laplace V_i^\text{cl}(x-y) \total^4 y \eqend{,} \\
h^\text{TT,qu}_{ij} &= 0 \eqend{.}
\end{align}
\end{subequations}
Inserting the solutions for the classical potentials~\eqref{spinning_efe_sol_classical} [or alternatively~\eqref{spinning_efe_svt_classical}] into the right-hand side, this further simplifies to
\begin{subequations}
\label{spinning_efe_sol_quantum}
\begin{align}
\Phi_\text{A}^\text{qu} &= \frac{\kappa^2 M}{6} \int \left[ 4 L_{C^2}(s,\vec{x}) + 3 L_{R^2}(s,\vec{x}) \right] \total s \eqend{,} \\
\Phi_\text{H}^\text{qu} &= \frac{\kappa^2 M}{6} \int \left[ 2 L_{C^2}(s,\vec{x}) - 3 L_{R^2}(s,\vec{x}) \right] \total s \eqend{,} \\
V_i^\text{qu} &= \kappa^2 \epsilon_{ijk} S^j \partial^k \int L_{C^2}(s,\vec{x}) \total s \eqend{,} \\
h^\text{TT,qu}_{ij} &= 0 \eqend{.}
\end{align}
\end{subequations}
To obtain expressions for the quantum corrections, it thus remains to calculate the Weyl and Ricci kernels for the different prescriptions contained in $L_{C^2/R^2}$~\eqref{calculation_kernel_l_def}, and integrate the resulting expressions over time, which we will do in the next section.

\section{The Weyl and Ricci kernels}
\label{sec_kernels}

In this section, we calculate the kernels $K^{AB}_{C^2/R^2}(x-y)$ for the ``$++$'' and ``$+-$'' prescriptions. We emphasise again that these two kernels are nothing else but the spin-2 and spin-0 parts of the graviton self-energy, which for free fields was calculated long ago in the time-ordered (the ``$++$'') case~\cite{capperduff1974,capperduffhalpern1974,capper1974}. As explained before~\eqref{calculation_kernels_func}, the corresponding result for the ``$+-$'' prescription can be simply obtained by Fourier transforming to coordinate space, and replacing the ``$++$'' prescription for $(x-y)^2$ by the ``$+-$'' prescription~\eqref{calculation_x2_prescriptions}. Moreover, for conformal theories (such as gauge fields in four dimensions, massless fermions or massless conformally coupled scalars), even strongly coupled ones, one could also use the general result for the two-point function of the stress-energy tensor~\cite{osbornshore2000}, which up to constant factors again gives exactly the kernels we need. However, we would like to present a way of calculation for massive quantum fields using Mellin-Barnes integrals, which works directly in coordinate space, and has the advantage that the results are both suited for numerical evaluation and allow a straightforward derivation of asymptotic expansions, both for small and large distances from the particle. Moreover, Mellin-Barnes integrals have been successfully used for calculations in (Anti-)de~Sitter space, where Mellin space seems to play the same simplifying role as Fourier space for a flat background~\cite{mack2009,penedones2011,fitzpatricketal2011,marolfmorrison2011,hollands2013,marolfmorrisonsrednicki2013,koraitanaka2013}, such that this calculation should be quite directly generalisable to those backgrounds.

\subsection{Gauge field}

It is well known that the classical action
\begin{equation}
S_0 \equiv - \frac{1}{4} \int F^{\mu\nu} F_{\mu\nu} \sqrt{-g} \total^n x
\end{equation}
with the field strength tensor
\begin{equation}
F_{\mu\nu} \equiv \nabla_\mu A_\nu - \nabla_\nu A_\mu
\end{equation}
constructed from the spin-1 field $A_\mu$ cannot be quantised straightforwardly because of gauge invariance, namely invariance of the action under the transformation
\begin{equation}
A_\mu \to A_\mu + \nabla_\mu \chi
\end{equation}
for an arbitrary function $\chi$. The modern way to deal with this gauge invariance is the BRST formalism~\cite{becchirouetstora1975,becchirouetstora1976,kugoojima1978,weinberg_v2,barnichbrandthenneaux2000}. One first introduces the usual ghost $c$, antighost $\bar{c}$ and auxiliary (Nakanishi-Lautrup) field $B$ in the theory, and then defines a differential $\brst$ by its action on the fields
\begin{equation}
\label{kernels_brst_action}
\brst A_\mu = \partial_\mu c \eqend{,} \qquad \brst c = 0 \eqend{,} \qquad \brst \bar{c} = \mathi B \eqend{,} \qquad \brst B = 0 \eqend{.}
\end{equation}
Furthermore, one defines $\brst$ to be fermionic, such that it satisfies a graded Leibniz rule
\begin{equation}
\brst (FG) = (\brst F) G \pm F (\brst G)
\end{equation}
for arbitrary functionals $F$ and $G$, with the sign depending on whether $F$ is bosonic or fermionic. From the explicit action~\eqref{kernels_brst_action} one also sees that the BRST differential is nilpotent $\brst^2 = 0$, and increases the ghost number by $1$ if one assigns ghost number $0$ to $A_\mu$ and $B$, ghost number $1$ to $c$ and ghost number $-1$ to $\bar{c}$. Gauge-fixing and ghost terms are then obtained by adding a term of the form $\brst \Psi$ to the action, where $\Psi$ is a suitable integrated functional of ghost number $-1$. For the usual covariant gauges, we take
\begin{equation}
\Psi = - \mathi \int \bar{c} \left( \frac{\xi}{2} B + G[A] \right) \sqrt{-g} \total^n x \eqend{,}
\end{equation}
with the gauge-fixing functional
\begin{equation}
G[A] \equiv \nabla_\mu A^\mu \eqend{,}
\end{equation}
and performing the BRST transformation the total action reads
\begin{equation}
\label{kernels_gauge_action}
\begin{split}
S &\equiv S_0 + \brst \Psi = - \frac{1}{4} \int F^{\mu\nu} F_{\mu\nu} \sqrt{-g} \total^n x - \frac{1}{2 \xi} \int \left( \nabla_\mu A^\mu \right)^2 \sqrt{-g} \total^n x \\
&\hspace{8em}+ \frac{1}{2 \xi} \int \left( \xi B + \nabla_\mu A^\mu \right)^2 \sqrt{-g} \total^n x + \mathi \int \bar{c} \nabla^2 c \sqrt{-g} \total^n x \eqend{.}
\end{split}
\end{equation}
Since the original action was gauge-invariant and the BRST transformation just acts as a gauge transformation with the gauge parameter replaced by the ghost~\eqref{kernels_brst_action}, we have $\brst F_{\mu\nu} = 0$, and since furthermore $\brst^2 = 0$ one sees that the gauge-fixed action is still BRST-invariant, $\brst S = 0$.

The advantage of this formalism is that one can see easily by a short calculation that the expectation value of a BRST-exact functional vanishes. Namely, one has
\begin{equation}
\label{kernels_brst_expectation}
\expect{\brst F}_\phi = \frac{\int \left( \brst F \right) \mathe^{\mathi S} \mathcal{D} \phi}{\int \mathe^{\mathi S} \mathcal{D} \phi} = \frac{\int \brst \left( F \mathe^{\mathi S} \right) \mathcal{D} \phi}{\int \mathe^{\mathi S} \mathcal{D} \phi} \eqend{,}
\end{equation}
for any functional $F$, where $\mathcal{D} \phi$ denotes an integral over all fields $A_\mu$, $c$, $\bar{c}$ and $B$, and where the second equality follows because of the BRST invariance of the total action. Now we have
\begin{equation}
\brst F = \pm \left( \partial_\mu c \right) \frac{\delta}{\delta A_\mu} F \pm \mathi B \frac{\delta}{\delta \bar{c}} F = \pm \frac{\delta}{\delta A_\mu} \left[ \left( \partial_\mu c \right) F \right] \pm \frac{\delta}{\delta \bar{c}} \left( \mathi B F \right)
\end{equation}
for any functional $F$ (with the signs depending on whether $F$ is bosonic or fermionic), and thus the integral in the numerator of equation~\eqref{kernels_brst_expectation} is a total derivative, and vanishes. In the same way, it is seen that expectation values of BRST-invariant functionals are independent of the choice of gauge-fixing functional $G[A]$, and more generally independent of $\Psi$: under the change $\Psi \to \Psi + \delta \Psi$ and for any functional $F$ with $\brst F = 0$, we have to first order in $\delta \Psi$
\begin{equation}
\int F \mathe^{\mathi \left( S + \brst \delta \Psi \right)} \mathcal{D} \phi = \int F \mathe^{\mathi S} \left( 1 + \mathi \brst \delta \Psi \right) \mathcal{D} \phi = \int F \mathe^{\mathi S} \mathcal{D} \phi \pm \mathi \int \brst \left( F \mathe^{\mathi S} \delta \Psi \right) \mathcal{D} \phi = \int F \mathe^{\mathi S} \mathcal{D} \phi \eqend{,}
\end{equation}
where the sign again depends on whether $F$ is bosonic or fermionic. In particular, classically gauge-invariant functionals are BRST-invariant, and their correlation functions are thus independent of the gauge fixing. These considerations are of course formal and dependent on a regulator which leaves the BRST transformations~\eqref{kernels_brst_action} unchanged, such as dimensional regularisation. However, one can (with much more effort) make them mathematically precise; see, e.g., Refs.~\cite{duetschfredenhagen2004,hollands2008,rejzner2011,fredenhagenrejzner2013,rejzner2015,froebhollandhollands2015} for a rigorous treatment of all IR, UV and gauge issues.

In particular, the stress tensor $T^{\mu\nu}$ defined by
\begin{equation}
T^{\mu\nu} \equiv - 2 \frac{\delta S}{\delta g^{\mu\nu}} = - 2 \frac{\delta S_0}{\delta g^{\mu\nu}} - 2 \brst \frac{\delta \Psi}{\delta g^{\mu\nu}} \equiv T_0^{\mu\nu} + \brst T_\Psi^{\mu\nu}
\end{equation}
is gauge- and thus BRST-invariant. Its two-point function, from which the Weyl and Ricci kernels are calculated according to equations~\eqref{calculation_tmunu_2pf} and~\eqref{calculation_kernel_k_def}, is thus independent of the gauge fixing, and moreover we have
\begin{equation}
\label{kernels_gauge_tmunu_t0munu}
\expect{ T^{\mu\nu}(x) T^{\rho\sigma}(y) }_\phi - \expect{ T^{\mu\nu}(x) }_\phi \expect{ T^{\rho\sigma}(y) }_\phi = \expect{ T_0^{\mu\nu}(x) T_0^{\rho\sigma}(y) }_\phi - \expect{ T_0^{\mu\nu}(x) }_\phi \expect{ T_0^{\rho\sigma}(y) }_\phi
\end{equation}
according to the general arguments presented above. While for Abelian theories (and thus in the free-field case) the ghosts decouple, and one can ignore them in purely gauge-theoretic calculations, the inclusion of their stress-energy is crucial for the equality~\eqref{kernels_gauge_tmunu_t0munu} to hold, since both gauge-fixing and ghost terms are generated from the same $\Psi$. Namely, if one were to perform an explicit calculation of the stress-tensor two-point function including $T_\Psi^{\mu\nu}$, one would find that the contribution from the ghosts exactly cancels the one from the gauge-fixing term, while the second-to-last term in the total action~\eqref{kernels_gauge_action} is algebraic and only gives rise to contact terms $\sim \delta^n(x-y)$, which can be absorbed in counterterms.

A short calculation using the expansions from Appendix~\ref{appendix_metric} leads for the flat Minkowski background to the well-known
\begin{equation}
T_0^{\mu\nu} = F^{\mu\alpha} F^\nu{}_\alpha - \frac{1}{4} \eta^{\mu\nu} F^{\alpha\beta} F_{\alpha\beta}
\end{equation}
and thus
\begin{equation}
\label{kernels_gauge_t0munu_in_f}
\begin{split}
&\expect{ T_0^{\mu\nu}(x) T_0^{\rho\sigma}(y) }_\phi - \expect{ T_0^{\mu\nu}(x) }_\phi \expect{ T_0^{\rho\sigma}(y) }_\phi = \mathcal{F}^{\mu\alpha\rho\gamma}(x,y) \mathcal{F}^\nu{}_\alpha{}^\sigma{}_\gamma(x,y) + \mathcal{F}^{\mu\alpha\sigma\gamma}(x,y) \mathcal{F}^\nu{}_\alpha{}^\rho{}_\gamma(x,y) \\
&\qquad\quad- \frac{1}{2} \eta^{\rho\sigma} \mathcal{F}^{\mu\alpha\gamma\delta}(x,y) \mathcal{F}^\nu{}_{\alpha\gamma\delta}(x,y) - \frac{1}{2} \eta^{\mu\nu} \mathcal{F}^{\alpha\beta\rho\gamma}(x,y) \mathcal{F}_{\alpha\beta}{}^\sigma{}_\gamma(x,y) \\
&\qquad\quad+ \frac{1}{8} \eta^{\mu\nu} \eta^{\rho\sigma} \mathcal{F}^{\alpha\beta\gamma\delta}(x,y) \mathcal{F}_{\alpha\beta\gamma\delta}(x,y)
\end{split}
\end{equation}
with
\begin{equation}
\mathcal{F}_{\mu\nu\rho\sigma}(x,y) \equiv \expect{ F_{\mu\nu}(x) F_{\rho\sigma}(y) }_\phi - \expect{ F_{\mu\nu}(x) }_\phi \expect{ F_{\rho\sigma}(y) }_\phi \eqend{.}
\end{equation}
This last expectation value can be evaluated using the gauge field two-point function
\begin{equation}
G_{\mu\nu}(x,y) \equiv - \mathi \expect{ A_\mu(x) A_\nu(y) }_\phi \eqend{,}
\end{equation}
which in turn is obtained from the quadratic part of the action~\eqref{kernels_gauge_action}. By shifting the auxiliary field $B \to B - \xi^{-1} \partial_\mu A^\mu$, only the first two terms in the action~\eqref{kernels_gauge_action} contribute, and we obtain
\begin{equation}
G_{\mu\nu}(x,y) = \eta_{\mu\nu} G_0((x-y)^2) - (1-\xi) \frac{\partial_\mu \partial_\nu}{\partial^2} G_0((x-y)^2) \eqend{,}
\end{equation}
where
\begin{equation}
\label{kernels_massless_propagator}
G_0(x^2) \equiv - \mathi \frac{\Gamma\left( \frac{n}{2}-1 \right)}{4 \pi^\frac{n}{2}} (x^2)^{-\frac{n-2}{2}}
\end{equation}
is the massless scalar field two-point function in $n$ dimensions, and the second term involving $\partial^{-2}$ can be calculated explicitly using
\begin{equation}
\label{kernels_x2_dalembert}
(x^2)^{-p} = \frac{1}{2(1-p)(n-2p)} \partial^2 (x^2)^{1-p}
\end{equation}
with $p = (n-2)/2$. We then calculate
\begin{equation}
\begin{split}
\mathcal{F}_{\mu\nu\rho\sigma}(x,y) &= - 2 \mathi \partial_\rho \partial_{[\mu} G_{\nu]\sigma}(x,y) + 2 \mathi \partial_\sigma \partial_{[\mu} G_{\nu]\rho}(x,y) \\
&= - 8 \mathi \eta_{\mu[\rho} \eta_{\sigma]\nu} G_0'((x-y)^2) + 16 \mathi (x-y)_{[\mu} \eta_{\nu][\rho} (x-y)_{\sigma]} G_0''((x-y)^2) \\
&= - 8 \mathi \left[ \eta_{\mu[\rho} \eta_{\sigma]\nu} + n \frac{(x-y)_{[\mu} \eta_{\nu][\rho} (x-y)_{\sigma]}}{(x-y)^2} \right] G_0'((x-y)^2) \\
\end{split}
\end{equation}
using that
\begin{equation}
x^2 G_0''(x^2) = - \frac{n}{2} G_0'(x^2) \eqend{,}
\end{equation}
as follows from the explicit expression~\eqref{kernels_massless_propagator} for the massless scalar two-point function. From equation~\eqref{kernels_gauge_t0munu_in_f} we then obtain
\begin{equation}
\label{kernels_gauge_tmunu2pf}
\begin{split}
&\expect{ T_0^{\mu\nu}(x) T_0^{\rho\sigma}(0) }_\phi - \expect{ T_0^{\mu\nu}(x) }_\phi \expect{ T_0^{\rho\sigma}(0) }_\phi = - 8 (n^2-8) \left( \eta^{\mu(\rho} \eta^{\sigma)\nu} - \eta^{\mu\nu} \eta^{\rho\sigma} \right) \left[ G_0'(x^2) \right]^2 \\
&\quad- 2 n (n-2) (n-1) \eta^{\mu\nu} \eta^{\rho\sigma} \left[ G_0'(x^2) \right]^2 + 16 n (3n-8) \frac{x^{(\mu} \eta^{\nu)(\rho} x^{\sigma)}}{x^2} \left[ G_0'(x^2) \right]^2 \\
&\quad+ 4 n (n-4)^2 \left( \eta^{\mu\nu} \frac{x^\rho x^\sigma}{x^2} + \eta^{\rho\sigma} \frac{x^\mu x^\nu}{x^2} \right) \left[ G_0'(x^2) \right]^2 - 8 n^2 (n-2) \frac{x^\mu x^\nu x^\rho x^\sigma}{(x^2)^2} \left[ G_0'(x^2) \right]^2 \eqend{.}
\end{split}
\end{equation}
where we have set $y = 0$ to shorten the expressions, since the two-point function is translation invariant.

Using the explicit form of the massless scalar two-point function~\eqref{kernels_massless_propagator}, one checks in a long but straightforward calculation that the connected stress tensor two-point function~\eqref{kernels_gauge_tmunu2pf} has the form~\eqref{calculation_tmunu_2pf}, where
\begin{subequations}
\begin{align}
f_1(x) &= \frac{(n^3-8n^2+10n+16)}{(n+1)(n-1)} \frac{\Gamma^2\left( \frac{n}{2}-1 \right)}{128 \pi^n} (x^2)^{2-n} \eqend{,} \\
f_2(x) &= \frac{(2n^2-3n-8)}{(n+1)(n-1)} \frac{\Gamma^2\left( \frac{n}{2}-1 \right)}{128 \pi^n} (x^2)^{2-n} \eqend{.}
\end{align}
\end{subequations}
The bare, unrenormalised kernels $K^\text{bare}_{C^2/R^2}$~\eqref{calculation_kernel_k_def} are thus given by
\begin{subequations}
\label{kernels_gauge_bare}
\begin{align}
K^\text{bare}_{C^2}(x) &= \mathi \frac{(2n^2-3n-8)(n-2)}{(n+1)(n-1)(n-3)} \frac{\Gamma^2\left( \frac{n}{2}-1 \right)}{512 \pi^n} (x^2)^{2-n} \eqend{,} \\
K^\text{bare}_{R^2}(x) &= \mathi \frac{(n-4)^2 (n-2)}{(n-1)^2} \frac{\Gamma^2\left( \frac{n}{2}-1 \right)}{1024 \pi^n} (x^2)^{2-n} \eqend{.}
\end{align}
\end{subequations}
For the ``$+-$'' prescription~\eqref{calculation_x2_prescription_pm}, i.e., the Wightman two-point function, $(x_{+-}^2)^{-2}$ is a well-defined distribution in four dimensions, and we can thus simply take the limit $n \to 4$ of the bare kernels. For the ``$++$'' prescription~\eqref{calculation_x2_prescription_pp}, i.e., the time-ordered two-point function, this is not the case. To extract the divergent part and obtain a renormalised kernel, we use equation~\eqref{kernels_x2_dalembert} with $p = n-2$ and add an ``intelligent zero'' to obtain
\begin{equation}
\label{kernels_gauge_x2_decomp}
(x^2)^{2-n} = \frac{1}{2(n-3)(n-4)} \partial^2 \left[ (x^2)^{3-n} - \mu^\frac{n-4}{2} (x^2)^{1-\frac{n}{2}} \right] + \frac{\mu^\frac{n-4}{2}}{2(n-3)(n-4)} \partial^2 (x^2)^{1-\frac{n}{2}}
\end{equation}
with the renormalisation scale $\mu$, introduced to make the above equation dimensionally correct. The first term has a well-defined limit as $n \to 4$, given by
\begin{equation}
\frac{1}{2(n-3)(n-4)} \partial^2 \left[ (x^2)^{3-n} - \mu^\frac{n-4}{2} (x^2)^{1-\frac{n}{2}} \right] \to - \frac{1}{4} \partial^2 \frac{\ln (\mu^2 x^2)}{x^2} \eqend{,}
\end{equation}
which for any prescription is a well-defined distribution in four dimensions, while using the massless scalar two-point function~\eqref{kernels_massless_propagator} the second term can be expressed as
\begin{equation}
\frac{\mu^\frac{n-4}{2}}{2(n-3)(n-4)} \partial^2 (x^2)^{1-\frac{n}{2}} = \mathi \frac{2 \pi^\frac{n}{2} \mu^\frac{n-4}{2}}{(n-3) (n-4) \Gamma\left( \frac{n}{2}-1 \right)} \partial^2 G_0(x^2) \eqend{.}
\end{equation}
Since for the ``$++$'' prescription $G_0(x_{++}^2)$ is the time-ordered two-point function, i.e., the propagator, we have
\begin{equation}
\partial^2 G_0(x_{++}^2) = \delta^n(x) \eqend{,}
\end{equation}
and thus this second term must be subtracted for the kernel $K^\text{bare}_{C^2}$ using the counterterm $\delta \alpha$~\eqref{calculation_sct}, while the explicit factor of $(n-4)^2$ in $K^\text{bare}_{R^2}$~\eqref{kernels_gauge_bare} leads to a vanishing contribution to $\delta \beta$ in the limit $n \to 4$. Since for the ``$+-$'' prescription $G_0(x_{+-}^2)$ is the Wightman function fulfilling $\partial^2 G_0(x_{+-}^2) = 0$, as explained before equation~\eqref{calculation_kernels_func} the renormalised kernels can be written in unified form
\begin{subequations}
\label{kernels_gauge_renormalised}
\begin{align}
K_{C^2}(x) &= - \frac{\mathi}{1280 \pi^4} \partial^2 \frac{\ln (\mu^2 x^2)}{x^2} \eqend{,} \\
K_{R^2}(x) &= 0 \eqend{,}
\end{align}
\end{subequations}
where the ``$++$'' and ``$+-$'' prescriptions are simply to be applied to $x^2$ according to equation~\eqref{calculation_x2_prescriptions}. This procedure is just the usual renormalisation, but performed in position space instead of the more well-known momentum space; see, e.g., Ref.~\cite{graciabondiagutierrezgarrovarilly2014} and references therein for more information.

The kernels $L_{C^2/R^2}$ appearing in the final expression for the quantum corrections to the gravitational potentials~\eqref{spinning_efe_sol_quantum} and defined by equation~\eqref{calculation_kernel_l_def} now read
\begin{subequations}
\label{kernels_gauge_l}
\begin{align}
L_{C^2}(x) &= - \frac{\mathi}{1280 \pi^4} \partial^2 \left[ \frac{\ln (\mu^2 x_{++}^2)}{x_{++}^2} - \frac{\ln (\mu^2 x_{+-}^2)}{x_{+-}^2} \right] + \alpha \delta^4(x) \eqend{,} \\
L_{R^2}(x) &= \beta \delta^4(x) \eqend{.}
\end{align}
\end{subequations}
The integral over time is calculated in Appendix~\ref{appendix_master} and given by equation~\eqref{appendix_master_result_log}, from which we finally obtain (with $r \equiv \abs{\vec{x}}$)
\begin{subequations}
\label{kernels_gauge_l_int}
\begin{align}
\int L_{C^2}(s,\vec{x}) \total s &= - \frac{1}{640 \pi^3} \laplace \frac{\ln(2 \mu r)}{r} + \alpha \delta^3(\vec{x}) = - \frac{1}{640 \pi^3} \laplace \frac{\ln(r)}{r} + \left[ \alpha + \frac{\ln(2 \mu)}{160 \pi^2} \right] \delta^3(\vec{x}) \eqend{,} \\
\int L_{R^2}(s,\vec{x}) \total s &= \beta \delta^3(\vec{x}) \eqend{,}
\end{align}
\end{subequations}
where the second equality was obtained using equation~\eqref{spinning_delta_laplace}. Note that we cannot evaluate $\laplace \ln(r)/r$ directly, since the result would be too singular at the origin to be a well-defined distribution. Only if we restrict to $r > 0$, we can calculate
\begin{equation}
\label{kernels_gauge_laplace_lnr_r}
\laplace \frac{\ln(r)}{r} = - \frac{1}{r^3} \qquad (r > 0) \eqend{,}
\end{equation}
and then of course the terms $\sim \delta^3(\vec{x})$ do not contribute either.

\subsection{Massive, minimally coupled scalar}

The most general action for a free scalar field $\phi$ is given by
\begin{equation}
S = - \frac{1}{2} \int \left( \nabla^\mu \phi \nabla_\mu \phi + m^2 \phi^2 + \xi R \phi^2 \right) \sqrt{-g} \total^n x \eqend{,}
\end{equation}
and includes a coupling to the Ricci curvature scalar with strength $\xi$. Using the expansions from Appendix~\ref{appendix_metric} and specializing to flat space, the corresponding stress tensor is easily calculated and reads
\begin{equation}
T^{\mu\nu} = \partial^\mu \phi \partial^\nu \phi - \frac{1}{2} \eta^{\mu\nu} \left( \partial^\rho \phi \partial_\rho \phi + m^2 \phi^2 \right) - \xi S^{\mu\nu} \phi^2 \eqend{.}
\end{equation}
For the case of minimal coupling $\xi = 0$, the renormalised stress-tensor two-point function has been calculated in position space in Ref.~\cite{froeb2013}. It is of the general form given in equation~\eqref{calculation_tmunu_2pf}, and the kernels $K^\text{bare}_{C^2/R^2}$ defined according to equation~\eqref{calculation_kernel_k_def} can be renormalised to obtain an expression of the form~\eqref{calculation_kernels_func}, where the renormalised kernels read
\begin{subequations}
\label{kernels_scalar_mc_renormalised}
\begin{align}
K_{C^2}(x) &= - \mathi \partial^2 \left( \frac{\ln(\mu^2 x^2)}{15360 \pi^4 x^2} \right) + \mathi \int_{\mathcal{C}^*} (m^2)^z (x^2)^{z-2} \frac{\Gamma(-z) \Gamma(1-z) \Gamma(2-z)}{2048 \pi^\frac{7}{2} \Gamma\left( \frac{7}{2} - z \right)} \frac{\total z}{2\pi\mathi} \eqend{,} \\
K_{R^2}(x) &= - \mathi \partial^2 \left( \frac{\ln(\mu^2 x^2)}{9216 \pi^4 x^2} \right) + \mathi \int_{\mathcal{C}^*} (m^2)^z (x^2)^{z-2} \frac{\Gamma(-z) \Gamma(2-z) \left[ 3 \Gamma(3-z) - \Gamma(1-z) \right]}{6144 \pi^\frac{7}{2} \Gamma\left( \frac{7}{2} - z \right)} \frac{\total z}{2\pi\mathi} \eqend{.}
\end{align}
\end{subequations}
The integrals appearing here are of Mellin-Barnes form, running over the contour $\mathcal{C}^*$ in the complex plane from $\Im z = - \mathi \infty$ to $\Im z = + \mathi \infty$ with $0 < \Re z < 1$ (see Ref.~\cite{froeb2013} for a short introduction to Mellin-Barnes integrals). Since the $\Gamma$ functions decay exponentially in imaginary directions~\cite{dlmf}, these integrals are absolutely convergent and well suited for numerical evaluation. It will be advantageous to further simplify the above expressions, and we use equation~\eqref{kernels_x2_dalembert} with $p = 2-z$ to extract a d'Alembertian operator from the integral (which is justified because of the absolute convergence, and since both before and after the extraction the integrals are well-defined distributions in four dimensions). Using $\Gamma$ function identities~\cite{dlmf} to simplify the integrands, this results in
\begin{subequations}
\begin{align}
K_{C^2}(x) &= - \mathi \partial^2 \left[ \frac{\ln(\mu^2 x^2)}{15360 \pi^4 x^2} - \int_{\mathcal{C}^*} (m^2)^z (x^2)^{z-1} \frac{\Gamma^2(-z) \Gamma(1-z)}{8192 \pi^\frac{7}{2} \Gamma\left( \frac{7}{2} - z \right)} \frac{\total z}{2\pi\mathi} \right] \eqend{,} \\
K_{R^2}(x) &= - \mathi \partial^2 \left[ \frac{\ln(\mu^2 x^2)}{9216 \pi^4 x^2} - \int_{\mathcal{C}^*} (m^2)^z (x^2)^{z-1} \frac{\Gamma^2(-z) \left[ 3 \Gamma(3-z) - \Gamma(1-z) \right]}{24576 \pi^\frac{7}{2} \Gamma\left( \frac{7}{2} - z \right)} \frac{\total z}{2\pi\mathi} \right] \eqend{.}
\end{align}
\end{subequations}
Define now the contour $\mathcal{C}$ to also run from $\Im z = - \mathi \infty$ to $\Im z = + \mathi \infty$, but with $-1 < \Re z < 0$. Since the integrands have only one pole between the two contours at $z = 0$ and are otherwise holomorphic, by the Cauchy integral and residue theorems we have
\begin{equation}
\label{kernels_scalar_mc_contour}
\int_{\mathcal{C}^*} f(z) \frac{\total z}{2\pi\mathi} = \int_\mathcal{C} f(z) \frac{\total z}{2\pi\mathi} + \operatorname{Res}_{z = 0} f(z) \eqend{,}
\end{equation}
and it follows that
\begin{subequations}
\label{kernels_scalar_mc}
\begin{align}
K_{C^2}(x) &= \mathi \partial^2 \left[ \frac{2 \gamma + \frac{46}{15} + \ln \left( \frac{m^2}{4 \mu^2} \right)}{15360 \pi^4 x^2} + \int_\mathcal{C} (m^2)^z (x^2)^{z-1} \frac{\Gamma^2(-z) \Gamma(1-z)}{8192 \pi^\frac{7}{2} \Gamma\left( \frac{7}{2} - z \right)} \frac{\total z}{2\pi\mathi} \right] \label{kernels_scalar_mc_kc2} \eqend{,} \\
K_{R^2}(x) &= \mathi \partial^2 \left[ \frac{2 \gamma + \frac{19}{15} + \ln \left( \frac{m^2}{4 \mu^2} \right)}{9216 \pi^4 x^2} + \int_\mathcal{C} (m^2)^z (x^2)^{z-1} \frac{\Gamma^2(-z) \left[ 3 \Gamma(3-z) - \Gamma(1-z) \right]}{24576 \pi^\frac{7}{2} \Gamma\left( \frac{7}{2} - z \right)} \frac{\total z}{2\pi\mathi} \right] \label{kernels_scalar_mc_kr2} \eqend{.}
\end{align}
\end{subequations}
Using the massless scalar two-point function~\eqref{kernels_massless_propagator}, we can express the first term as
\begin{equation}
\frac{1}{x^2} = 4 \pi^2 \mathi G_0(x^2) \eqend{.}
\end{equation}
Since for the ``$+-$'' prescription $G_0(x_{+-}^2)$ is the (negative) Wightman function, we have $\partial^2 G_0(x_{+-}^2) = 0$ and the first terms drop out of the kernels~\eqref{kernels_scalar_mc}. For the ``$++$'' prescription, however, $G_0(x_{++}^2)$ is the propagator and we have
\begin{equation}
\partial^2 G_0(x_{++}^2) = \delta^4(x) \eqend{.}
\end{equation}
These terms can then be absorbed by a finite renormalisation of the parameters $\alpha$ and $\beta$ in the effective action~\eqref{calculation_seff_inin_ren} [or alternatively in equation~\eqref{calculation_kernel_l_def}], and we will assume that this has been done, such that the kernels $K_{C^2/R^2}$ only consist of the integral terms in equation~\eqref{kernels_scalar_mc}.

It then only remains to calculate the integrals~\eqref{spinning_efe_sol_quantum} for the combinations $L_{C^2/R^2}$~\eqref{calculation_kernel_l_def}, which can be done using Appendix~\ref{appendix_master}, specifically the result~\eqref{appendix_master_result_z}, and again using the absolute convergence of the Mellin-Barnes integrals to justify the exchange of integrals. We then obtain (with $r \equiv \abs{\vec{x}}$)
\begin{subequations}
\begin{align}
\int L_{C^2}(s,\vec{x}) \total s &= \partial^2 \int_\mathcal{C} (m^2)^z r^{2z-1} \frac{\Gamma^2(-z) \Gamma\left( \frac{1}{2} - z \right)}{8192 \pi^3 \Gamma\left( \frac{7}{2} - z \right)} \frac{\total z}{2\pi\mathi} + \alpha \delta^3(\vec{x}) \eqend{,} \\
\int L_{R^2}(s,\vec{x}) \total s &= \partial^2 \int_\mathcal{C} (m^2)^z r^{2z-1} \frac{\Gamma^2(-z) \left[ 3 \Gamma(3-z) - \Gamma(1-z) \right] \Gamma\left( \frac{1}{2} - z \right)}{24576 \pi^3 \Gamma(1-z) \Gamma\left( \frac{7}{2} - z \right)} \frac{\total z}{2\pi\mathi} + \beta \delta^3(\vec{x}) \eqend{,}
\end{align}
\end{subequations}
and using that
\begin{equation}
\partial^2 r^{2z-1} = \laplace r^{2z-1}
\end{equation}
and some $\Gamma$ function identities~\cite{dlmf}, this simplifies to
\begin{subequations}
\label{kernels_scalar_mc_l_int}
\begin{align}
\int L_{C^2}(s,\vec{x}) \total s &= \laplace \int_\mathcal{C} (m^2)^z r^{2z-1} \frac{\Gamma^2(-z)}{1024 \pi^3 (1-2z) (3-2z) (5-2z)} \frac{\total z}{2\pi\mathi} + \alpha \delta^3(\vec{x}) \eqend{,} \\
\int L_{R^2}(s,\vec{x}) \total s &= \laplace \int_\mathcal{C} (m^2)^z r^{2z-1} \frac{\Gamma^2(-z) \left[ 3 (2-z) (1-z) - 1 \right]}{3072 \pi^3 (1-2z) (3-2z) (5-2z)} \frac{\total z}{2\pi\mathi} + \beta \delta^3(\vec{x}) \eqend{.}
\end{align}
\end{subequations}
Again, since we have $-1 < \Re z < 0$ on the integration contour $\mathcal{C}$, we cannot evaluate the Laplacian directly as the result would be too singular at the origin to be a well-defined distribution. If we restrict to $r > 0$ we have
\begin{equation}
\laplace r^{2z-1} = 2z (2z-1) r^{2z-3} \qquad (r > 0) \eqend{,}
\end{equation}
and then the local terms $\sim \delta^3(\vec{x})$ have to be disregarded as well.

\subsection{Massive scalar with general curvature coupling}

The fastest way to arrive at the proper expressions for $\xi \neq 0$ is to reuse the result of Mart{\'\i}n and Verdaguer~\cite{martinverdaguer2000}, who tell us that in the general case the kernel $K_{C^2}$ is $\xi$-independent [and thus equal to its value for $\xi = 0$~\eqref{kernels_scalar_mc_kc2}], while the kernel $K_{R^2}$ has a factor of
\begin{equation}
\left( (1-6\xi) + 2 \frac{m^2}{\partial^2} \right)^2
\end{equation}
acting on a $\xi$-independent function. We thus have to rewrite our result~\eqref{kernels_scalar_mc_kr2}, which has $\xi = 0$, to include a factor of $\left( 1 + 2 m^2 \partial^{-2} \right)^2$, and can then simply perform the extension (it has been checked in Ref.~\cite{froeb2013} that the Fourier transform of the result~\eqref{kernels_scalar_mc_kr2} coincides with the minimal-coupling result of Ref.~\cite{martinverdaguer2000}).

For this, we first calculate (using equation~\eqref{kernels_x2_dalembert} and shifting the integration variable)
\begin{equation}
\label{kernels_scalar_gc_shift}
m^2 \partial^{-2} \int_\mathcal{C} (m^2)^z (x^2)^{z-1} f(z) \frac{\total z}{2\pi\mathi} = \int_\mathcal{C} (m^2)^z (x^2)^{z-1} \frac{1}{4 z (z-1)} f(z-1) \frac{\total z}{2\pi\mathi} \eqend{,}
\end{equation}
such that
\begin{equation}
\begin{split}
&\left( 1 + 2 \frac{m^2}{\partial^2} \right)^2 \int_\mathcal{C} (m^2)^z (x^2)^{z-1} f(z) \frac{\total z}{2\pi\mathi} \\
&\quad= \int_\mathcal{C} (m^2)^z (x^2)^{z-1} \left[ f(z) + \frac{f(z-1)}{z (z-1)} + \frac{f(z-2)}{4 z (z-1)^2 (z-2)} \right] \frac{\total z}{2\pi\mathi} \eqend{.}
\end{split}
\end{equation}
Comparing with the kernel $K_{R^2}$ for the minimally coupled case~\eqref{kernels_scalar_mc_kr2}, we thus have to find a function $f(z)$ such that
\begin{equation}
f(z) + \frac{f(z-1)}{z (z-1)} + \frac{f(z-2)}{4 z (z-1)^2 (z-2)} = \frac{\Gamma^2(-z) \left[ 3 \Gamma(3-z) - \Gamma(1-z) \right]}{\Gamma\left( \frac{7}{2} - z \right)} \eqend{,}
\end{equation}
which a bit of guesswork reveals to be
\begin{equation}
f(z) = \frac{4 \Gamma(1-z) \Gamma^2(-z)}{3 \Gamma\left( \frac{3}{2} - z \right)} \eqend{.}
\end{equation}
The kernel $K_{R^2}$ for the minimally coupled case~\eqref{kernels_scalar_mc_kr2} can thus be written as
\begin{equation}
K_{R^2}(x) = \mathi \partial^2 \left( 1 + 2 \frac{m^2}{\partial^2} \right)^2 \int_\mathcal{C} (m^2)^z (x^2)^{z-1} \frac{\Gamma(1-z) \Gamma^2(-z)}{18432 \pi^\frac{7}{2} \Gamma\left( \frac{3}{2} - z \right)} \frac{\total z}{2\pi\mathi} \eqend{,}
\end{equation}
and the extension to general curvature coupling reads
\begin{equation}
\begin{split}
K_{R^2}(x) &= \mathi \partial^2 \left( 1-6\xi + 2 \frac{m^2}{\partial^2} \right)^2 \int_\mathcal{C} (m^2)^z (x^2)^{z-1} \frac{\Gamma(1-z) \Gamma^2(-z)}{18432 \pi^\frac{7}{2} \Gamma\left( \frac{3}{2} - z \right)} \frac{\total z}{2\pi\mathi} \\
&= \mathi \partial^2 \int_\mathcal{C} (m^2)^z (x^2)^{z-1} \frac{\Gamma^2(-z) \Gamma(1-z)}{73728 \pi^\frac{7}{2} \Gamma\left( \frac{7}{2} - z \right)} \\
&\qquad\times \left[ (1-6\xi)^2 (5-2z) (3-2z) - 2 (1-6\xi) (5-2z) z + z (z-1) \right] \frac{\total z}{2\pi\mathi} \eqend{,}
\end{split}
\end{equation}
where we used equation~\eqref{kernels_scalar_gc_shift} and $\Gamma$ function identities~\cite{dlmf} to arrive at the second equality.

The calculation of the integral~\eqref{spinning_efe_sol_quantum} for the combination $L_{R^2}$~\eqref{calculation_kernel_l_def} is now done in the same way as for the minimally coupled case, and we obtain (with $r \equiv \abs{\vec{x}}$)
\begin{equation}
\label{kernels_scalar_gc_l_int}
\begin{split}
\int L_{R^2}(s,\vec{x}) \total s &= \beta \delta^3(\vec{x}) + \laplace \int_\mathcal{C} (m^2)^z r^{2z-1} \frac{\Gamma^2(-z)}{9216 \pi^3 (5-2z) (3-2z) (1-2z)} \\
&\qquad\times \left[ (1-6\xi)^2 (5-2z) (3-2z) - 2 (1-6\xi) (5-2z) z + z (z-1) \right] \frac{\total z}{2\pi\mathi} \eqend{.}
\end{split}
\end{equation}

\subsection{Massive fermion}

For the $\gamma$ matrices and the spin connection in curved space, we follow the conventions of Weinberg~\cite{weinberg_v1} and Freedman/van Proeyen~\cite{freedmanvanproeyen}, to which we refer the reader for details (with the main difference to usual particle physics texts being the absence of most factors of $\mathi$). The action for a free massive fermion reads
\begin{equation}
- \int \bar{\psi} \left( \gamma^\mu \nabla_\mu - m \right) \psi \total^n x \eqend{,}
\end{equation}
and the (symmetric) stress tensor in a flat-space background is given by
\begin{equation}
T_{\mu\nu} = \frac{1}{2} \bar{\psi} \gamma_{(\mu} \partial_{\nu)} \psi - \frac{1}{2} \left( \partial_{(\nu} \bar{\psi} \right) \gamma_{\mu)} \psi \eqend{.}
\end{equation}
The fermionic propagator $\mathcal{G}_m(x)$ can be obtained from the massive scalar propagator $G_{m^2}(x^2)$ in the usual way
\begin{equation}
\mathcal{G}_m(x) \equiv - \mathi \expect{ \psi(x) \bar{\psi}(0) } = - \left( \gamma^\mu \partial_\mu + m \right) G_{m^2}(x^2) \eqend{.}
\end{equation}

For the stress tensor two-point function we then obtain
\begin{equation}
\label{kernels_fermion_tmunu_prop}
\begin{split}
\mathcal{T}_{\mu\nu\rho\sigma}(x,y) &\equiv \expect{ T_{\mu\nu}(x) T_{\rho\sigma}(y) }_\phi - \expect{ T_{\mu\nu}(x) }_\phi \expect{ T_{\rho\sigma}(y) }_\phi \\
&= \frac{1}{4} \tr \left[ \gamma_{(\mu} \partial^x_{\nu)} \mathcal{G}_m(x-y) \gamma_{(\rho} \partial^y_{\sigma)} \mathcal{G}_m(y-x) \right] \\
&\quad- \frac{1}{4} \tr \left[ \gamma_{(\mu} \partial^x_{\nu)} \partial^y_{(\rho} \mathcal{G}_m(x-y) \gamma_{\sigma)} \mathcal{G}_m(y-x) \right] \\
&\quad- \frac{1}{4} \tr \left[ \mathcal{G}_m(x-y) \gamma_{(\rho} \partial^y_{\sigma)} \partial^x_{(\mu} \mathcal{G}_m(y-x) \gamma_{\nu)} \right] \\
&\quad+ \frac{1}{4} \tr \left[ \partial^y_{(\rho} \mathcal{G}_m(x-y) \gamma_{\sigma)} \partial^x_{(\mu} \mathcal{G}_m(y-x) \gamma_{\nu)} \right] \\
&= 2 \eta_{\mu(\rho} \eta_{\sigma)\nu} G'_{m^2}((x-y)^2) \left[ - n G'_{m^2}((x-y)^2) + m^2 G_{m^2}((x-y)^2) \right] \tr \unitmatrix \\
&\quad+ 2 \eta_{\mu\nu} \eta_{\rho\sigma} G'_{m^2}((x-y)^2) G'_{m^2}((x-y)^2) \tr \unitmatrix \\
&\quad- 16 x_{(\mu} \eta_{\nu)(\rho} x_{\sigma)} G'_{m^2}((x-y)^2) G''_{m^2}((x-y)^2) \tr \unitmatrix \\
&\quad+ 16 x_\mu x_\nu x_\rho x_\sigma \left[ G''_{m^2}((x-y)^2) G''_{m^2}((x-y)^2) - G'_{m^2}((x-y)^2) G'''_{m^2}((x-y)^2) \right] \tr \unitmatrix \\
\end{split}
\end{equation}
where we have used the usual ($n$-dimensional) $\gamma$ matrix algebra to evaluate the matrix trace $\tr$, and $\tr \unitmatrix$ is the dimension of the representation, equal to $4$ in $n = 4$ dimensions. To put this into the general form~\eqref{calculation_tmunu_2pf} and perform renormalisation, we use the following Mellin-Barnes integral representation from~\cite{froeb2013} (note that there a factor of $\mathi$ was removed from the definition of $G$, which leads to an additional minus sign in comparison)
\begin{equation}
\label{kernels_fermion_propagator_product}
G^{(k)}_{m^2}(x^2) G^{(l)}_{m^2}(x^2) = - \int_\mathcal{C} (m^2)^z (x^2)^{z+2-k-l-n} \frac{(-1)^{k+l}}{4^{2+z} \pi^n} K(k,l,z) \frac{\total z}{2\pi\mathi}
\end{equation}
with
\begin{equation}
\label{kernels_fermion_k_def}
K(k,l,z) = \frac{\Gamma(n-2+k+l-z) \Gamma\left( \frac{n}{2}-1+k-z \right) \Gamma\left( \frac{n}{2}-1+l-z \right) \Gamma(-z)}{\Gamma(n-2+k+l-2z)} \eqend{,}
\end{equation}
where the contour $\mathcal{C}$ runs from $\Im z = - \mathi \infty$ to $\Im z = + \mathi \infty$ left of all poles of $K(k,l,z)$. By translation invariance, we can set $y = 0$, and the stress tensor two-point function~\eqref{kernels_fermion_tmunu_prop} can then be written in Mellin-Barnes form
\begin{equation}
\label{kernels_fermion_tmunu_mb}
\begin{split}
&\mathcal{T}_{\mu\nu\rho\sigma}(x,0) = \int_\mathcal{C} (m^2)^z (x^2)^{z-n} \frac{1}{4^{1+z} \pi^n} \bigg[ 2 \eta_{\mu(\rho} \eta_{\sigma)\nu} \left( n K(1,1,z) + 4 K(1,0,z-1) \right) \\
&\ - 2 \eta_{\mu\nu} \eta_{\rho\sigma} K(1,1,z) - 16 \frac{x_{(\mu} \eta_{\nu)(\rho} x_{\sigma)}}{x^2} K(1,2,z) - 16 \frac{x_\mu x_\nu x_\rho x_\sigma}{(x^2)^2} \left( K(2,2,z) - K(1,3,z) \right) \bigg] \frac{\total z}{2\pi\mathi} \eqend{.}
\end{split}
\end{equation}

To bring this into the general form~\eqref{calculation_tmunu_2pf}, we make an ansatz using the $S_{\mu\nu}$ operators~\eqref{calculation_smunu_def} of the form
\begin{equation}
\label{kernels_fermion_tmunu_s}
\mathcal{T}_{\mu\nu\rho\sigma}(x,0) = \int_\mathcal{C} (m^2)^z \left[ f(z) S_{\mu\nu} S_{\rho\sigma} + g(z) S_{\mu(\rho} S_{\sigma)\nu} \right] (x^2)^{z+2-n} \frac{\Gamma(n-z) \Gamma^2\left( \frac{n}{2}-z \right) \Gamma(-z)}{4^{1+z} \pi^n \Gamma(n+2-2z)} \frac{\total z}{2\pi\mathi} \eqend{,}
\end{equation}
and performing the derivatives and comparing with~\eqref{kernels_fermion_tmunu_mb} it follows that
\begin{subequations}
\begin{align}
f(z) &= - \frac{(n-2z)}{2 (z+2-n) (z+1-n) (n-2-2z)} \eqend{,} \\
g(z) &= - (n-1-2z) f(z) \eqend{.}
\end{align}
\end{subequations}
The left-most pole of the integrand is located at $z = 0$, and we can thus take the contour $\mathcal{C}$ to be at $-1 < \Re z < 0$, just as for the scalar case. However, $(x^2)^{z+2-n}$ is not a well-defined distribution in $n = 4$ dimensions for $\Re z < 0$, and we thus have to shift the contour to $\Re z > 0$. This can be done using equation~\eqref{kernels_scalar_mc_contour}, and we pick up an additional term given by the residue of the integrand at $z = 0$. This term is now proportional to $(x^2)^{2-n}$, which can be renormalised in the same way as for the gauge field [compare equation~\eqref{kernels_gauge_x2_decomp} and the following discussion]. The remaining Mellin-Barnes integral over the new contour $\mathcal{C}^*$ is now a well-defined distribution, and we can take the limit $n \to 4$ there. Similar to the scalar case, we can finally extract a d'Alembertian operator from this integral, and shift the contour back to $-1 < \Re z < 0$ to obtain a simple renormalised expression, possible performing an additional finite renormalisation [compare equation~\eqref{kernels_scalar_mc_renormalised} and the following discussion]. Since we are not interested in the details of the renormalisation, and just need the final renormalised expression, we can simply extract a d'Alembertian operator from the integral~\eqref{kernels_fermion_tmunu_s} using equation~\eqref{kernels_x2_dalembert} with $p = n-2-z$ and take the limit $n \to 4$ of the resulting expression. Using some $\Gamma$ function identities~\cite{dlmf}, this gives
\begin{equation}
\mathcal{T}_{\mu\nu\rho\sigma}(x,0) = \partial^2 \int_\mathcal{C} (m^2)^z \left[ - S_{\mu\nu} S_{\rho\sigma} + (3-2z) S_{\mu(\rho} S_{\sigma)\nu} \right] (x^2)^{z-1} \frac{\Gamma(1-z) \Gamma^2(-z)}{1024 \pi^\frac{7}{2} \Gamma\left( \frac{7}{2} - z \right)} \frac{\total z}{2\pi\mathi} \eqend{,}
\end{equation}
and the renormalised kernels $K_{C^2/R^2}$~\eqref{calculation_kernels_func} can be calculated by comparing this result with equations~\eqref{calculation_tmunu_2pf} and~\eqref{calculation_kernel_k_def} and read
\begin{subequations}
\begin{align}
K_{C^2}(x) &= \mathi \partial^2 \int_\mathcal{C} (m^2)^z (x^2)^{z-1} \frac{\Gamma(1-z) \Gamma^2(-z) (3-2z)}{4096 \pi^\frac{7}{2} \Gamma\left( \frac{7}{2} - z \right)} \frac{\total z}{2\pi\mathi} \eqend{,} \\
K_{R^2}(x) &= \mathi \partial^2 \int_\mathcal{C} (m^2)^z (x^2)^{z-1} \frac{\Gamma^2(1-z) \Gamma(-z)}{12288 \pi^\frac{7}{2} \Gamma\left( \frac{7}{2} - z \right)} \frac{\total z}{2\pi\mathi} \eqend{.}
\end{align}
\end{subequations}

The calculation of the integrals~\eqref{spinning_efe_sol_quantum} for the combinations $L_{C^2/R^2}$~\eqref{calculation_kernel_l_def} is now done in the same way as for the scalar case, using the integral~\eqref{appendix_master_result_z} calculated in Appendix~\ref{appendix_master}, and we obtain (with $r \equiv \abs{\vec{x}}$, and using some $\Gamma$ function identities~\cite{dlmf})
\begin{equation}
\label{kernels_fermion_l_int}
\begin{split}
\int L_{C^2}(s,\vec{x}) \total s &= \laplace \int_\mathcal{C} (m^2)^z r^{2z-1} \frac{\Gamma^2(-z)}{512 \pi^3 (1-2z) (5-2z)} \frac{\total z}{2\pi\mathi} + \alpha \delta^3(\vec{x}) \eqend{,} \\
\int L_{R^2}(s,\vec{x}) \total s &= \laplace \int_\mathcal{C} (m^2)^z r^{2z-1} \frac{\Gamma(1-z) \Gamma(-z)}{1536 \pi^3 (1-2z) (3-2z) (5-2z)} \frac{\total z}{2\pi\mathi} + \beta \delta^3(\vec{x}) \eqend{.}
\end{split}
\end{equation}

\section{Results}
\label{sec_results}

Since for very small distances $r$ from the particle, the test particle approximation that we use breaks down (since there the particle's own gravitational field is strong and we cannot neglect the backreaction anymore), we can restrict to $r > 0$ when presenting the results. We can then evaluate the Laplacians acting on the expressions~\eqref{kernels_gauge_l_int}, \eqref{kernels_scalar_mc_l_int}, \eqref{kernels_scalar_gc_l_int} and~\eqref{kernels_fermion_l_int}, and the local terms appearing in these results do not contribute.

Combining the classical~\eqref{spinning_efe_sol_classical} and quantum contributions~\eqref{spinning_efe_sol_quantum} to the gravitational potentials according to equation~\eqref{calculation_efe_classical_quantum_decomp}, we have
\begin{subequations}
\begin{align}
\Phi_\text{A} &= \frac{\kappa^2 M}{16 \pi r} \left[ 1 + \frac{8 \pi \kappa^2 r}{3} \int \left[ 4 L_{C^2}(s,\vec{x}) + 3 L_{R^2}(s,\vec{x}) \right] \total s \right] \eqend{,} \\
\Phi_\text{H} &= \frac{\kappa^2 M}{16 \pi r} \left[ 1 + \frac{8 \pi \kappa^2 r}{3} \int \left[ 2 L_{C^2}(s,\vec{x}) - 3 L_{R^2}(s,\vec{x}) \right] \total s \right] \eqend{,} \\
V_i &= - \frac{\kappa^2 (\vec{S} \times \vec{r})_i}{8 \pi r^3} \left[ 1 - 8 \pi \kappa^2 r^2 \partial_r \int L_{C^2}(s,\vec{x}) \total s \right] \eqend{.}
\end{align}
\end{subequations}
For the gauge field~\eqref{kernels_gauge_l_int}, this gives
\begin{subequations}
\label{result_gauge}
\begin{align}
\Phi_\text{A} &= \frac{\kappa^2 M}{16 \pi r} \left[ 1 + \frac{\kappa^2}{60 \pi^2 r^2} \right] \eqend{,} \\
\Phi_\text{H} &= \frac{\kappa^2 M}{16 \pi r} \left[ 1 + \frac{\kappa^2}{120 \pi^2 r^2} \right] \eqend{,} \\
V_i &= - \frac{\kappa^2 (\vec{S} \times \vec{r})_i}{8 \pi r^3} \left[ 1 + \frac{3 \kappa^2}{80 \pi^2 r^2} \right] \eqend{,}
\end{align}
\end{subequations}
for the massive scalar with general curvature coupling~\eqref{kernels_scalar_mc_l_int},~\eqref{kernels_scalar_gc_l_int} we obtain
\begin{subequations}
\label{result_scalar}
\begin{align}
\Phi_\text{A} &= \frac{\kappa^2 M}{16 \pi r} \left[ 1 + \frac{\kappa^2 [ 1 + \frac{5}{4} (1-6\xi)^2 ]}{720 \pi^2 r^2} \int_\mathcal{C} (m r)^{2z} f_\text{A}(z) \frac{\total z}{2\pi\mathi} \right] \eqend{,} \label{result_scalar_phia} \\
\Phi_\text{H} &= \frac{\kappa^2 M}{16 \pi r} \left[ 1 + \frac{\kappa^2 [ 1 - \frac{5}{2} (1-6\xi)^2 ]}{1440 \pi^2 r^2} \int_\mathcal{C} (m r)^{2z} f_\text{H}(z) \frac{\total z}{2\pi\mathi} \right] \eqend{,} \\
V_i &= - \frac{\kappa^2 (\vec{S} \times \vec{r})_i}{8 \pi r^3} \left[ 1 + \frac{\kappa^2}{320 \pi^2 r^2} \int_\mathcal{C} (m r)^{2z} \frac{5 \Gamma(1-z) \Gamma(-z)}{(5-2z)} \frac{\total z}{2\pi\mathi} \right] \eqend{,}
\end{align}
\end{subequations}
with
\begin{subequations}
\begin{align}
f_\text{A}(z) &\equiv \frac{5 \Gamma(1-z) \Gamma(-z)}{5 (1-6\xi)^2 + 4} \left[ (1-6\xi)^2 - \frac{2 (1-6\xi) z}{(3-2z)} + \frac{z (z-1) + 12}{(3-2z) (5-2z)} \right] \eqend{,} \\
f_\text{H}(z) &\equiv \frac{5 \Gamma(1-z) \Gamma(-z)}{5 (1-6\xi)^2 - 2} \left[ (1-6\xi)^2 - \frac{2 (1-6\xi) z}{(3-2z)} + \frac{z (z-1) - 6}{(3-2z) (5-2z)} \right] \eqend{,}
\end{align}
\end{subequations}
and for a massive fermion~\eqref{kernels_fermion_l_int} we get
\begin{subequations}
\label{result_fermion}
\begin{align}
\Phi_\text{A} &= \frac{\kappa^2 M}{16 \pi r} \left[ 1 + \frac{\kappa^2}{120 \pi^2 r^2} \int_\mathcal{C} (m r)^{2z} \frac{15 \Gamma(1-z) \Gamma(-z) (4-3z)}{4 (3-2z) (5-2z)} \frac{\total z}{2\pi\mathi} \right] \eqend{,} \label{result_fermion_phia} \\
\Phi_\text{H} &= \frac{\kappa^2 M}{16 \pi r} \left[ 1 + \frac{\kappa^2}{240 \pi^2 r^2} \int_\mathcal{C} (m r)^{2z} \frac{15 \Gamma(1-z) \Gamma(-z) (2-z)}{2 (3-2z) (5-2z)} \frac{\total z}{2\pi\mathi} \right] \eqend{,} \\
V_i &= - \frac{\kappa^2 (\vec{S} \times \vec{r})_i}{8 \pi r^3} \left[ 1 + \frac{3 \kappa^2}{160 \pi^2 r^2} \int_\mathcal{C} (m r)^{2z} \frac{5 \Gamma(1-z) \Gamma(-z) (3-2z)}{3 (5-2z)} \frac{\total z}{2\pi\mathi} \right] \eqend{.}
\end{align}
\end{subequations}
These are the main results of this article, which we now discuss in more detail.

\subsection{Small and zero masses}
\label{sec_results_smallzeromass}

\begin{figure}
\begin{minipage}[b]{.47\linewidth}
\centering\includegraphics[width=\textwidth]{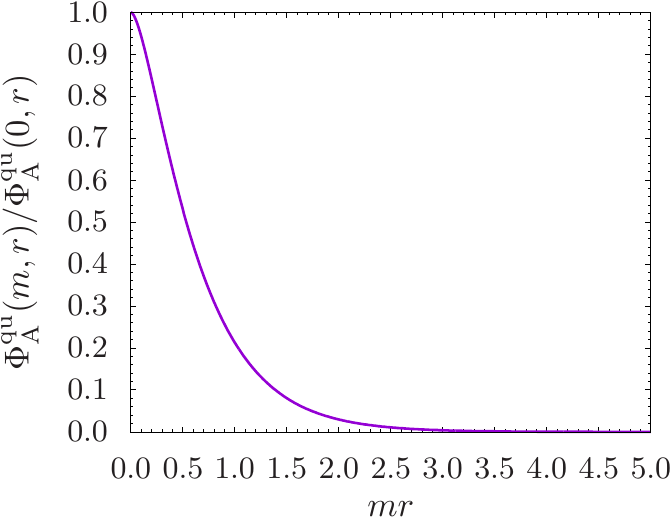}
\subcaption{\parbox[t]{.9\linewidth}{Quantum corrections to the scalar-type potential $\Phi_\text{A}$ due to fermions.\\\phantom{with $\xi = 1/6$.}}}
\end{minipage}\hfil
\begin{minipage}[b]{.47\linewidth}
\centering\includegraphics[width=\textwidth]{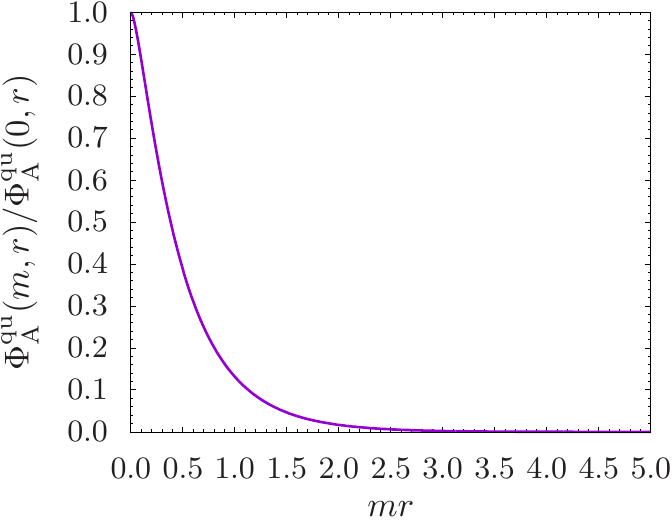}
\subcaption{\parbox[t]{.9\linewidth}{Quantum corrections to the scalar-type potential $\Phi_\text{A}$ due to conformally coupled scalars with $\xi = 1/6$.}}
\end{minipage}\\[1em]
\begin{minipage}[b]{.47\linewidth}
\centering\includegraphics[width=\textwidth]{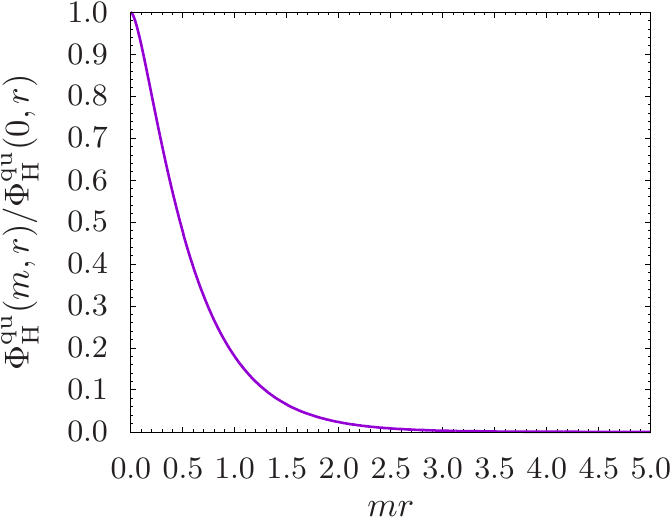}
\subcaption{\parbox[t]{.9\linewidth}{Quantum corrections to the scalar-type potential $\Phi_\text{H}$ due to fermions.\\\phantom{with $\xi = 1/6$.}}}
\end{minipage}\hfil
\begin{minipage}[b]{.47\linewidth}
\centering\includegraphics[width=\textwidth]{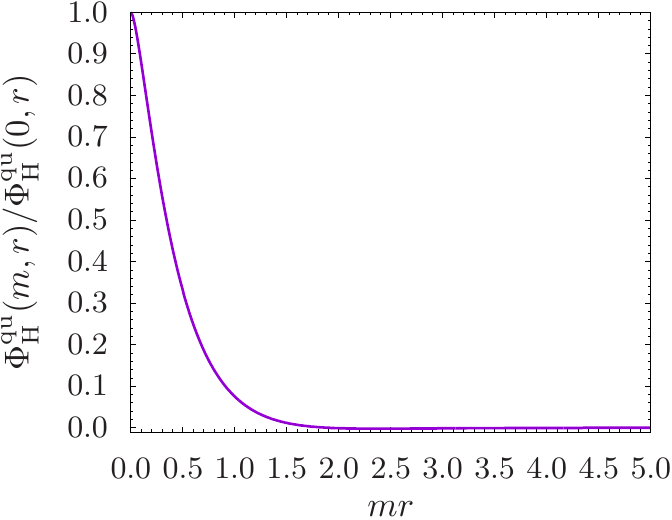}
\subcaption{\parbox[t]{.9\linewidth}{Quantum corrections to the scalar-type potential $\Phi_\text{H}$ due to conformally coupled scalars with $\xi = 1/6$.}}
\end{minipage}\\[1em]
\begin{minipage}[b]{.47\linewidth}
\centering\includegraphics[width=\textwidth]{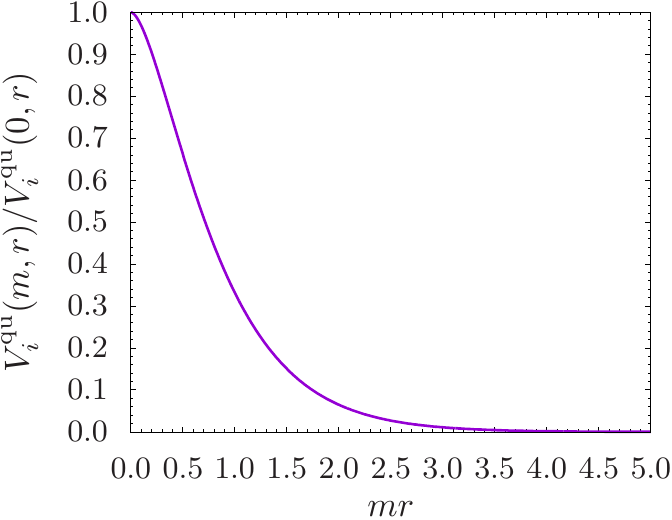}
\subcaption{\parbox[t]{.9\linewidth}{Quantum corrections to the vector-type potential $V_i$ due to fermions.}}
\end{minipage}\hfil
\begin{minipage}[b]{.47\linewidth}
\centering\includegraphics[width=\textwidth]{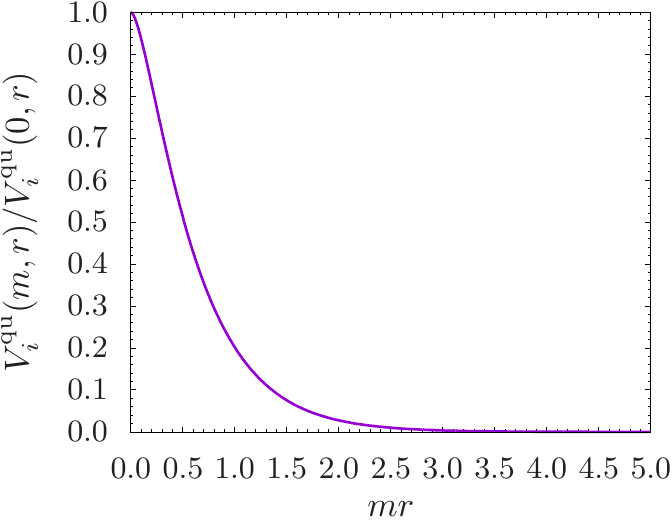}
\subcaption{\parbox[t]{.9\linewidth}{Quantum corrections to the vector-type potential $V_i$ due to scalars.}}
\end{minipage}
\caption{Quantum corrections to the gravitational potentials due to scalars and fermions of mass $m$ in comparison to the massless case.}\label{fig_corr_1}
\end{figure}

\begin{figure}
\begin{minipage}[b]{.47\linewidth}
\centering\includegraphics[width=\textwidth]{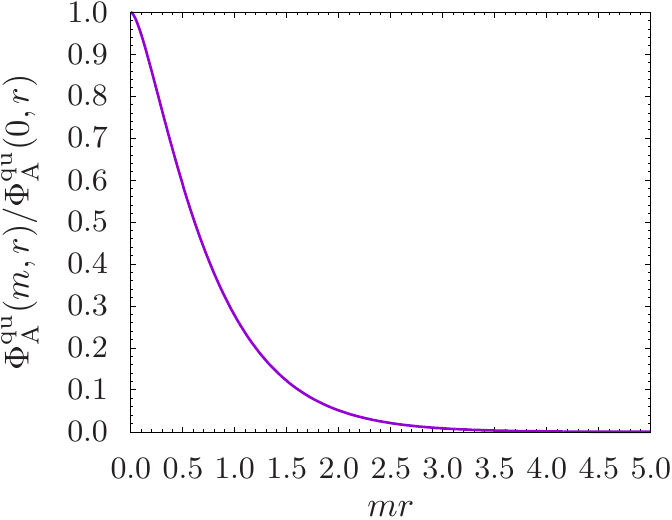}
\subcaption{\parbox[t]{.9\linewidth}{Quantum corrections to the scalar-type potential $\Phi_\text{A}$ due to minimally coupled scalars with $\xi = 0$.}}
\end{minipage}\hfil
\begin{minipage}[b]{.47\linewidth}
\centering\includegraphics[width=\textwidth]{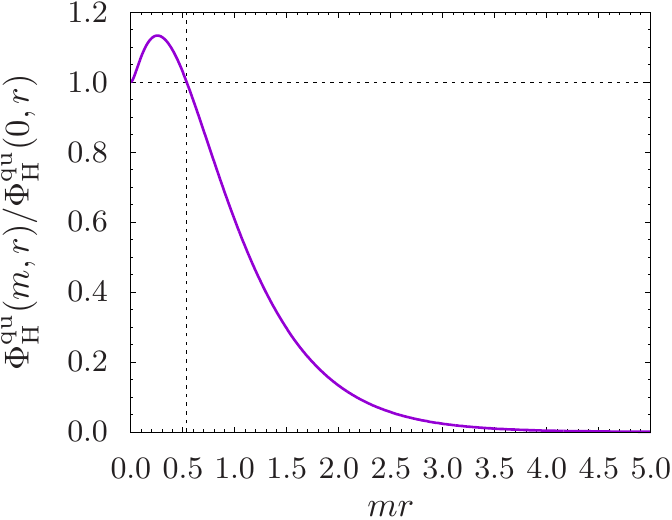}
\subcaption{\parbox[t]{.9\linewidth}{Quantum corrections to the scalar-type potential $\Phi_\text{H}$ due to minimally coupled scalars with $\xi = 0$.}}\label{fig_corr_2_phih_xi0}
\end{minipage}\\[1em]
\begin{minipage}[b]{.47\linewidth}
\centering\includegraphics[width=\textwidth]{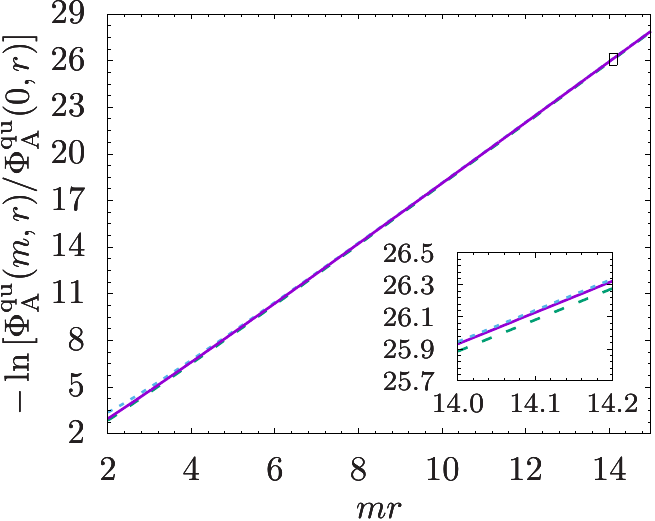}
\subcaption{\parbox[t]{.9\linewidth}{Asymptotic form of the quantum corrections to the scalar-type potential $\Phi_\text{A}$ due to minimally coupled scalars with $\xi = 0$. The solid violet line is the numerical data, the dashed green one is the first-order asymptotic expansion and the dotted blue one is the second-order asymptotic expansion.}}
\label{fig_corr_2_phia_asymp}
\end{minipage}\hfil
\begin{minipage}[b]{.47\linewidth}
\centering\includegraphics[width=\textwidth]{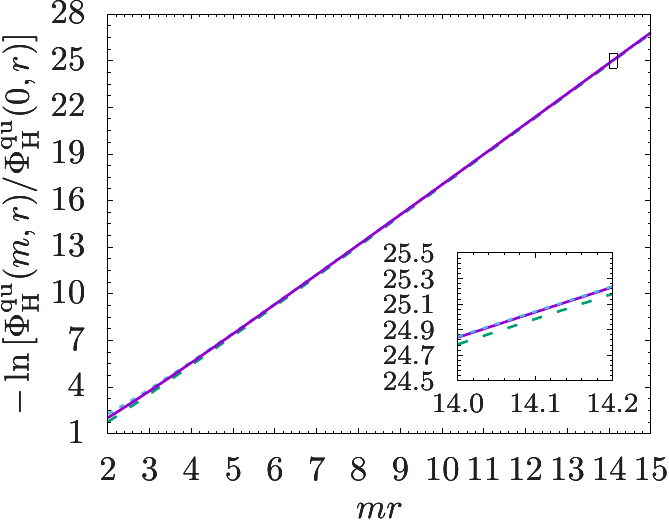}
\subcaption{\parbox[t]{.9\linewidth}{Asymptotic form of the quantum corrections to the scalar-type potential $\Phi_\text{H}$ due to minimally coupled scalars with $\xi = 0$. The solid violet line is the numerical data, the dashed green one is the first-order asymptotic expansion and the dotted blue one is the second-order asymptotic expansion.}}
\label{fig_corr_2_phih_asymp}
\end{minipage}
\caption{Quantum corrections to the gravitational potentials due to scalars and fermions of mass $m$ in comparison to the massless case (continued).}\label{fig_corr_2}
\end{figure}

Note first that the Mellin-Barnes integrals are normalised such that they equal $1$ for $m = 0$. In the massless case, we thus have
\begin{subequations}
\label{result_massless}
\begin{align}
\Phi_\text{A} &= \frac{\kappa^2 M}{16 \pi r} \left[ 1 + \left[ N_0 \left( 1 + \frac{5}{4} (1-6\xi)^2 \right) + 6 N_{1/2} + 12 N_1 \right] \frac{\kappa^2}{720 \pi^2 r^2} \right] \label{result_massless_phia} \eqend{,} \\
\Phi_\text{H} &= \frac{\kappa^2 M}{16 \pi r} \left[ 1 + \left[ N_0 \left( 1 - \frac{5}{2} (1-6\xi)^2 \right) + 6 N_{1/2} + 12 N_1 \right] \frac{\kappa^2}{1440 \pi^2 r^2} \right] \eqend{,} \\
V_i &= - \frac{\kappa^2 (\vec{S} \times \vec{r})_i}{8 \pi r^3} \left[ 1 + \left( N_0 + 6 N_{1/2} + 12 N_1 \right) \frac{\kappa^2}{320 \pi^2 r^2} \right] \eqend{,}
\end{align}
\end{subequations}
where $N_s$ is the number of spin-$s$ fields. Since in the nonrelativistic limit, $\Phi_\text{A}$ gives minus the Newtonian potential $V(r)$, and $\kappa^2 = 16 \pi G_\text{N}$, we have full agreement with the well-known existing result~\eqref{introduction_vr}. The interesting changes due to massive particles then reside in the integrals, i.e., in
\begin{equation}
\frac{\Phi_\text{A}^\text{qu}(m,r)}{\Phi_\text{A}^\text{qu}(0,r)} = \int_\mathcal{C} (m r)^{2z} f_\text{A}(z) \frac{\total z}{2\pi\mathi}
\end{equation}
(for the scalar case), and the corresponding other ratios of quantum corrections. Since the $\Gamma$ functions in the integrand fall off exponentially, they integrals are easily evaluated numerically, and the graphs are shown in figures~\ref{fig_corr_1} and~\ref{fig_corr_2}. As one can see from the figures, the corrections die off fast, and since the quantum corrections to the gravitational potentials are already tiny in the case of massless particles, these corrections are not accessible to experiment in any foreseeable future. Nevertheless, the quantum correction to the second Bardeen potential $\Phi_\text{H}^\text{qu}$ shows, for distances of the order of the Compton wavelength of the virtual particle, enhancement over the massless case for certain values of the non-minimal coupling parameter $\xi$ (e.g., for the minimally-coupled case $\xi = 0$, see figure~\ref{fig_corr_2_phih_xi0}), and in fact grows without bound for $\xi \to (1\pm\sqrt{2/5})/6$. One might thus think that this enhancement could have observable consequences, but it is just the value of $\xi$ for which the massless correction vanishes, and the full quantum correction $\Phi_\text{H}^\text{qu}$ stays tiny for all values of $\xi$.


For small masses in general, we can shift the contour $\mathcal{C}$ of the Mellin-Barnes integrals to the right, picking up residues from the poles that lie between the old and new contour. The integrals have a series of poles at integer $z$, coming from the $\Gamma$ functions in the numerator, and two isolated ones at $z = 3/2$ and $z = 5/2$. For example, taking the new contour $\mathcal{C}'$ to have $3/2 < \Re z < 2$, we have [analogously to equation~\eqref{kernels_scalar_mc_contour}]
\begin{equation}
\int_\mathcal{C} f(z) \frac{\total z}{2\pi\mathi} = \int_{\mathcal{C}'} f(z) \frac{\total z}{2\pi\mathi} - \sum_{z_i \in \left\{0,1,\frac{3}{2}\right\}} \operatorname{Res}_{z = z_i} f(z) \eqend{.}
\end{equation}
The integral over the contour $\mathcal{C}'$ is still absolutely convergent and we can bound it by a constant times $(mr)^{2 \Re z}$, and since we can shift the contour to have $\Re z$ as close to $2$ as we like and the pole at $z = 2$ is of order $2$, this is a term of order $\bigo{m^4 \ln m}$. For the scalar with general curvature coupling, we obtain in this way
\begin{subequations}
\label{result_scalar_smallm}
\begin{align}
\begin{split}
\frac{\Phi_\text{A}^\text{qu}(m,r)}{\Phi_\text{A}^\text{qu}(0,r)} &= 1 + \frac{10 [ 1+18\xi-18\xi^2 + 3 (1+12\xi^2) ( \ln(mr) + \gamma ) ]}{3 (3 - 20 \xi + 60 \xi^2)} m^2 r^2 \\
&\qquad- \frac{5 \pi (3+16\xi)}{2 (3 - 20 \xi + 60 \xi^2)} m^3 r^3 + \bigo{m^4 \ln m} \eqend{,}
\end{split} \label{result_scalar_smallm_phia} \\
\begin{split}
\frac{\Phi_\text{H}^\text{qu}(m,r)}{\Phi_\text{H}^\text{qu}(0,r)} &= 1 + \frac{10 [ -4+18\xi-18\xi^2 + 3 (-1+12\xi^2) ( \ln(mr) + \gamma ) ]}{3 (1 - 20 \xi + 60 \xi^2)} m^2 r^2 \\
&\qquad- \frac{5 \pi (-5+16\xi)}{2 (1 - 20 \xi + 60 \xi^2)} m^3 r^3 + \bigo{m^4 \ln m} \eqend{,}
\end{split} \\
\frac{V_i^\text{qu}(m,r)}{V_i^\text{qu}(0,r)} &= 1 + \frac{5 [ -1 + 6 ( \ln(mr) + \gamma ) ]}{9} m^2 r^2 + \bigo{m^4 \ln m} \eqend{,}
\end{align}
\end{subequations}
and for fermions, we get
\begin{subequations}
\label{result_fermion_smallm}
\begin{align}
\frac{\Phi_\text{A}^\text{qu}(m,r)}{\Phi_\text{A}^\text{qu}(0,r)} &= 1 + \frac{5 [ -2 + 3 ( \ln(mr) + \gamma ) ]}{6} m^2 r^2 + \frac{5\pi}{4} m^3 r^3 + \bigo{m^4 \ln m} \eqend{,} \label{result_fermion_smallm_phia} \\
\frac{\Phi_\text{H}^\text{qu}(m,r)}{\Phi_\text{H}^\text{qu}(0,r)} &= 1 + \frac{5 [ 1 + 3 ( \ln(mr) + \gamma ) ]}{3} m^2 r^2 - \frac{5\pi}{2} m^3 r^3 + \bigo{m^4 \ln m} \eqend{,} \\
\frac{V_i^\text{qu}(m,r)}{V_i^\text{qu}(0,r)} &= 1 + \frac{5 [ -7 + 6 ( \ln(mr) + \gamma ) ]}{27} m^2 r^2 + \bigo{m^4 \ln m} \eqend{.}
\end{align}
\end{subequations}

\subsection{Large masses and distances}

On the other hand, for large distances (and masses), we can shift the contour arbitrarily far to the left without changing the value of the integral, since there are no poles for $\Re z < 0$. Thus, the quantum corrections fall of faster than any power of $m$, and from the graphs one might suspect exponential decay. That this is in fact the case is shown in Appendix~\ref{appendix_asymptotic}, where also the explicit form of the asymptotic expansion is derived for a general Mellin-Barnes integral of the type we are considering. Using the integral $I_a(mr)$ defined in equation~\eqref{appendix_asymptotic_int} we have
\begin{subequations}
\label{result_scalar_asymptotic_ia}
\begin{align}
\frac{\Phi_\text{A}^\text{qu}(m,r)}{\Phi_\text{A}^\text{qu}(0,r)} &= \frac{45}{4 [ 5 (1-6\xi)^2 + 4 ]} \left[ (1+12\xi^2) I_1(m r) - 4 \xi (1-4\xi) I_2(m r) + (1-4\xi)^2 I_3(m r) \right] \eqend{,} \\
\frac{\Phi_\text{H}^\text{qu}(m,r)}{\Phi_\text{H}^\text{qu}(0,r)} &= \frac{45}{4 [ 5 (1-6\xi)^2 - 2 ]} \left[ - (1-12\xi^2) I_1(m r) - 4 \xi (1-4\xi) I_2(m r) + (1-4\xi)^2 I_3(m r) \right] \eqend{,} \\
\frac{V_i^\text{qu}(m,r)}{V_i^\text{qu}(0,r)} &= \frac{5}{4} I_1(m r) + \frac{5}{2} I_2(m r)
\end{align}
\end{subequations}
for a massive scalar with general curvature coupling, and
\begin{subequations}
\label{result_fermion_asymptotic_ia}
\begin{align}
\frac{\Phi_\text{A}^\text{qu}(m,r)}{\Phi_\text{A}^\text{qu}(0,r)} &= \frac{15}{16} I_1(m r) + \frac{45}{16} I_2(m r) \eqend{,} \\
\frac{\Phi_\text{H}^\text{qu}(m,r)}{\Phi_\text{H}^\text{qu}(0,r)} &= \frac{15}{8} I_1(m r) + \frac{15}{8} I_2(m r) \eqend{,} \\
\frac{V_i^\text{qu}(m,r)}{V_i^\text{qu}(0,r)} &= \frac{5}{12} I_1(m r) + \frac{5}{3} I_3(m r)
\end{align}
\end{subequations}
for a massive fermion. The asymptotic expansion of $I_a(m r)$ to next-to-leading order is given by equation~\eqref{appendix_asymptotic_expansion}, and we obtain
\begin{subequations}
\label{result_scalar_asymptotic}
\begin{align}
\frac{\Phi_\text{A}^\text{qu}(m,r)}{\Phi_\text{A}^\text{qu}(0,r)} &= \frac{45 (1-4\xi)^2}{4 [ 5 (1-6\xi)^2 + 4 ]} \sqrt{\pi} \, \mathe^{-2 m r} (m r)^\frac{1}{2} \left[ 1 - \frac{13+12\xi}{16 (1-4\xi) m r} + \bigo{ \frac{1}{m^2 r^2}} \right] \eqend{,} \label{result_scalar_asymptotic_phia} \\
\frac{\Phi_\text{H}^\text{qu}(m,r)}{\Phi_\text{H}^\text{qu}(0,r)} &= \frac{45 (1-4\xi)^2}{4 [ 5 (1-6\xi)^2 - 2 ]} \sqrt{\pi} \, \mathe^{-2 m r} (m r)^\frac{1}{2} \left[ 1 - \frac{13+12\xi}{16 (1-4\xi) m r} + \bigo{ \frac{1}{m^2 r^2}} \right] \eqend{,} \\
\frac{V_i^\text{qu}(m,r)}{V_i^\text{qu}(0,r)} &= \frac{5}{2} \sqrt{\pi} \, \mathe^{-2 m r} (m r)^{- \frac{1}{2}} \left[ 1 - \frac{41}{16 m r} + \bigo{ \frac{1}{m^2 r^2}} \right]
\end{align}
\end{subequations}
for a massive scalar with general curvature coupling, and
\begin{subequations}
\label{result_fermion_asymptotic}
\begin{align}
\frac{\Phi_\text{A}^\text{qu}(m,r)}{\Phi_\text{A}^\text{qu}(0,r)} &= \frac{45}{16} \sqrt{\pi} \, \mathe^{-2 m r} (m r)^{- \frac{1}{2}} \left[ 1 - \frac{131}{48 m r} + \bigo{ \frac{1}{m^2 r^2}} \right] \eqend{,} \label{result_fermion_asymptotic_phia} \\
\frac{\Phi_\text{H}^\text{qu}(m,r)}{\Phi_\text{H}^\text{qu}(0,r)} &= \frac{15}{8} \sqrt{\pi} \, \mathe^{-2 m r} (m r)^{- \frac{1}{2}} \left[ 1 - \frac{33}{16 m r} + \bigo{ \frac{1}{m^2 r^2}} \right] \eqend{,} \\
\frac{V_i^\text{qu}(m,r)}{V_i^\text{qu}(0,r)} &= \frac{5}{3} \sqrt{\pi} \, \mathe^{-2 m r} (m r)^\frac{1}{2} \left[ 1 - \frac{13}{16 m r} + \bigo{ \frac{1}{m^2 r^2}} \right]
\end{align}
\end{subequations}
for a massive fermion.

The asymptotic expansions to first and second order are plotted together with the numerical result for the Bardeen potentials for the minimally-coupled scalar in figures~\ref{fig_corr_2_phia_asymp} and~\ref{fig_corr_2_phih_asymp}. One can see that the approximations are extremely good already for small distances $r$ from the particle, and become virtually indistinguishable for large distances.

\subsection{Comparison with previous results}

Apart from few exceptions, existing calculations of quantum corrections only consider corrections to the Newtonian potential $V(r)$, to which the first Bardeen potential $\Phi_\text{A}$ reduces in the non-relativistic limit. Moreover, most of these calculations focus on the case of massless virtual particles, either matter fields (which we also treat in this work) or gravitons. As already stated in subsection~\ref{sec_results_smallzeromass}, our results in the massless case are in full agreement with the known ones for the Newtonian potential~\cite{radkowski1970,schwinger1968,duff1974,capperduffhalpern1974,capperduff1974,donoghue1994a,donoghue1994b,muzinichvokos1995,hamberliu1995,akhundovbelluccishiekh1997,duffliu2000a,duffliu2000b,kirilinkhriplovich2002,khriplovichkirilin2003,bjerrumbohrdonoghueholstein2003a,bjerrumbohrdonoghueholstein2003b,satzmazzitellialvarez2005,parkwoodard2010,marunovicprokopec2011,marunovicprokopec2012}. Refs.~\cite{satzmazzitellialvarez2005,parkwoodard2010,marunovicprokopec2011,marunovicprokopec2012} are also considering general quantum corrections to the metric due to loops of massless scalars, and their result reads (simplified and converted to our notation)
\begin{subequations}
\label{comparison_massless_scalar}
\begin{align}
h_{00} &= \frac{\kappa^4 M}{7680 \pi^3 r^3} \left( 3 - 20 \xi + 60 \xi^2 \right) \eqend{,} \\
h_{0i} &= 0 \eqend{,} \\
h_{ij} &= \frac{\kappa^4 M}{7680 \pi^3 r^3} \left( -1 + 20 \xi - 60 \xi^2 \right) \delta_{ij} \eqend{.}
\end{align}
\end{subequations}
Since these results were derived in an unknown gauge, we cannot directly compare them with our results for the gauge-invariant gravitational potentials. However, looking at the decompositions~\eqref{calculation_hmunu_inv_gauge} and~\eqref{calculation_hmunu_gaugeinvpart}, we see that
\begin{subequations}
\begin{align}
\Phi_\text{A} &= \frac{1}{2} h_{00} \eqend{,} \\
\Phi_\text{H} &= \frac{1}{4} \left( \delta^{ij} h_{ij} - \frac{\partial^i \partial^j}{\laplace} h_{ij} \right)
\end{align}
\end{subequations}
in any gauge where the metric perturbation does not explicitly depend on time, as for the results above. Thus, the result~\eqref{comparison_massless_scalar} gives
\begin{subequations}
\begin{align}
\Phi_\text{A} &= \frac{\kappa^4 M}{15360 \pi^3 r^3} \left( 3 - 20 \xi + 60 \xi^2 \right) \eqend{,} \\
\Phi_\text{H} &= \frac{\kappa^4 M}{15360 \pi^3 r^3} \left( -1 + 20 \xi - 60 \xi^2 \right) \eqend{,}
\end{align}
\end{subequations}
which coincides exactly with our result in the massless case for non-spinning particles~\eqref{result_massless}.

The only reference that presents explicit results for the Newtonian potential as a function of distance $r$ in the massive case seems to be the recent work of Burns and Pilaftsis~\cite{burnspilaftsis2015}, treating massive minimally coupled scalars, massive fermions and massive (Proca-type) vector bosons. Their general result for the quantum corrections to the Newtonian potential is given by the integrals
\begin{equation}
\Delta V(r) = \frac{G_\text{N}}{60 \pi} \int_{2m}^\infty \mathe^{-q r} \left( 3 - \frac{4m^2}{q^2} + \frac{28m^4}{q^4} \right) \sqrt{q^2 - 4 m^2} \total q
\end{equation}
for minimally coupled scalars~\cite{burnspilaftsis2015}, and
\begin{equation}
\Delta V(r) = \frac{G_\text{N}}{15 \pi} \int_{2m}^\infty \mathe^{-q r} \left( 2 - \frac{m^2}{q^2} - \frac{28m^4}{q^4} \right) \sqrt{q^2 - 4 m^2} \total q
\end{equation}
for fermions~\cite{burnspersonal}. While it hasn't been possible to bring our general result for $\Phi_\text{A}$~\eqref{result_scalar_phia} and~\eqref{result_fermion_phia} in this form, we can compare the small- and large-mass expansions. If those coincide, the simplicity of both our and their result then makes it highly probable that the full results coincide as well.

For small masses, Ref.~\cite{burnspilaftsis2015} obtains for the quantum corrections
\begin{equation}
\Delta V(r) = \frac{G_\text{N}}{20 \pi r^2} \left[ 1 + \frac{10}{3} m^2 r^2 \left[ \ln(mr) + \gamma + \frac{1}{3} \right] + \bigo{m^3 r^3} \right]
\end{equation}
for minimally coupled scalars and
\begin{equation}
\Delta V(r) = \frac{2 G_\text{N}}{15 \pi r^2} \left[ 1 + \frac{5}{2} m^2 r^2 \left[ \ln(mr) + \gamma - \frac{2}{3} \right] + \bigo{m^3 r^3} \right]
\end{equation}
for fermions (correcting a missing factor of $2$ for the massless case~\cite{burnspersonal}). Since the massless case already agrees with the known results~\eqref{introduction_vr}, we can simply compare the terms in brackets with the quotients~\eqref{result_scalar_smallm_phia} for the scalar case, setting $\xi = 0$ to obtain the minimally-coupled result, and~\eqref{result_fermion_smallm_phia} for fermions, and using that $\kappa^2 = 16 \pi G_\text{N}$ we find full agreement. For large masses, however, their expansion does not match with ours --- which might be due to the neglect of some subleading terms in the expansion of special functions~\cite{burnspersonal}, and can be rectified. Setting $x \equiv 2 m r$ and making the change of variables $q = 2m (t+1)$, their result reads
\begin{subequations}
\begin{align}
\Delta V(r) &= \frac{G_\text{N} m^2}{15 \pi} \mathe^{-x} \int_0^\infty \mathe^{- x t} \left( 3 - \frac{1}{(t+1)^2} + \frac{7}{4 (t+1)^4} \right) \sqrt{t^2+2t} \total t \eqend{,} \\
\Delta V(r) &= \frac{4 G_\text{N} m^2}{15 \pi} \mathe^{-x} \int_0^\infty \mathe^{- x t} \left( 2 - \frac{1}{4 (t+1)^2} - \frac{7}{4 (t+1)^4} \right) \sqrt{t^2+2t} \total t
\end{align}
\end{subequations}
for scalars and fermions, respectively. Both of the integrals are of the form
\begin{equation}
\int_0^\infty \mathe^{- x t} f(t) \total t
\end{equation}
with $f(t)$ having an asymptotic expansion of the type
\begin{equation}
f(t) \sim \sum_{s=0}^\infty a_s t^{s+\lambda-1}
\end{equation}
as $t \to 0$. In the scalar case, we have $\lambda = 3/2$ and
\begin{equation}
a_0 = \frac{15}{4} \sqrt{2} \eqend{,} \qquad a_1 = - \frac{65}{16} \sqrt{2} \eqend{,}
\end{equation}
while for fermions it results $\lambda = 5/2$ and
\begin{equation}
a_0 = \frac{15}{2} \sqrt{2} \eqend{,} \qquad a_1 = - \frac{131}{8} \sqrt{2} \eqend{.}
\end{equation}
By Watson's Lemma~\cite{olver}, the asymptotic expansion of the integral as $x \to \infty$ is then given by
\begin{equation}
\int_0^\infty \mathe^{- x t} f(t) \total t \sim \sum_{s=0}^\infty \Gamma(s+\lambda) \frac{a_s}{x^{s+\lambda}} \eqend{,}
\end{equation}
and we obtain
\begin{subequations}
\begin{align}
\Delta V(r) &\sim \frac{G m^2}{16 \sqrt{\pi}} \mathe^{-2 m r} (m r)^{-3/2} \left[ 1 - \frac{13}{16 m r} + \bigo{ \frac{1}{(m r)^2} } \right] \eqend{,} \\
\Delta V(r) &\sim \frac{3 G m^2}{8 \sqrt{\pi}} \mathe^{-2 m r} (m r)^{-5/2} \left[ 1 - \frac{131}{48 m r} + \bigo{ \frac{1}{(m r)^2} } \right]
\end{align}
\end{subequations}
for scalars and fermions, respectively. Combining the massless result~\eqref{result_massless_phia} with the large-mass expansions~\eqref{result_scalar_asymptotic_phia} for scalars (setting $\xi = 0$ to obtain the minimally-coupled case) and~\eqref{result_fermion_asymptotic_phia} for fermions, we again have full agreement between this expansion and our results. Thus, since both our and their result are given by quite simple integrals, it is highly probably that they fully coincide, even if it has not been possible to prove this directly.

All these comparisons have been for spinless particles, since as explained in the introduction our calculation is different from one the undertaken in Ref.~\cite{holsteinross2008}. Ref.~\cite{bjerrumbohrdonoghueholstein2003a} calculates quantum corrections to the metric perturbation for a spin-1/2 particle, but these corrections are due to virtual gravitons and not due to matter. Nevertheless, their results have the same form as ours in the massless case~\eqref{result_massless}, but with different numerical prefactors.

\section{Discussion}
\label{sec_discussion}

We have derived the corrections to the gauge-invariant gravitational potentials for spinning particles due to loops of massive and massless quantum fields. This includes the Newtonian potential, for which these corrections have been studied previously, and we have found full agreement with existing results. However, there is one more scalar-type potential for which only corrections due to massless fields have been studied, and a vector-type (gravitomagnetic) potential where those corrections were unexplored. Unfortunately, the results are too tiny to be measured experimentally in the foreseeable future, but they are important in principle, especially for providing unambiguous results for low-energy quantum gravitational predictions which must be reproduced in any full theory of quantum gravity.

The method by which we arrived at the results was quite different from the usual one, which is based on inferring a Newtonian potential from scattering data (the inverse scattering method). Instead, similar to how the classical Newtonian potential is obtained by solving the gravitational field equations for a point source, we have solved the field equations coming from an effective gravitational action, which includes loop corrections of massive particles. The main advantage of this method over the inverse scattering method is its applicability in curved spacetimes, where a scattering matrix may not be present. In fact, in these cases it seems to be the only method available. Although this paper did not deal with a curved background, but Minkowski spacetime, the calculation is still somewhat simpler than the corresponding one using the inverse scattering method~\cite{radkowski1970,schwinger1968,duff1974,capperduffhalpern1974,capperduff1974,donoghue1994a,donoghue1994b,muzinichvokos1995,hamberliu1995,akhundovbelluccishiekh1997,duffliu2000a,duffliu2000b,kirilinkhriplovich2002,khriplovichkirilin2003,bjerrumbohrdonoghueholstein2003a,bjerrumbohrdonoghueholstein2003b,satzmazzitellialvarez2005,parkwoodard2010,marunovicprokopec2011,marunovicprokopec2012}, and seems comparable in complexity to a recent calculation using modern techniques for scattering amplitudes~\cite{bjerrumbohretal2016}. In particular, the calculation of the effective action essentially boils down to the calculation of the graviton self-energy (including renormalisation), and we could simply have used the well-known results of Capper et al.~\cite{capperduffhalpern1974,capperduff1974,capper1974}. To obtain the Newtonian potential, and expansions both for small and large distances from the particle in coordinate space, we would then only have had to perform a Fourier transform of their momentum-space expression. However, the Mellin-Barnes integral representation we employed has several advantages: the results are well suited for numerical evaluation, and they allow a straightforward derivation of asymptotic expansions, both for small and large distances from the particle. Moreover, since Mellin-Barnes integrals have already been used successfully in (Anti-)de~Sitter space~\cite{mack2009,penedones2011,fitzpatricketal2011,marolfmorrison2011,hollands2013,marolfmorrisonsrednicki2013,koraitanaka2013}, our calculation should be quite immediately generalisable to those backgrounds.

Since the effective action is gauge invariant, and thus must be expressible using gauge-invariant variables only, our method provides a further non-trivial check on the correctness of the calculation. This has a further advantage in the case at hand: since the equations determining the Newtonian potential (and the other gravitational potentials) are constraint equations for the gauge-invariant variables, only a spatial Laplacian needs to be inverted, which gives an unambigously determined result for the quantum corrections~\eqref{spinning_efe_sol_quantum}, and no dynamical differential equation needs to be solved. Note, however, that at higher orders the definition of the Newtonian potential becomes ambiguous (see, e.g.,~\cite{bjerrumbohrdonoghueholstein2003a,bjerrumbohrdonoghueholstein2003b} and references therein), and this ambiguity will also show up using our method. Since the scattering matrix is gauge, and generally reparametrisation invariant~\cite{kalloshtyutin1972,lam1973}, the full scattering amplitude does not suffer from such ambiguities. Thus, the scattering amplitude seems to be preferable to characterise quantum gravitational corrections at higher orders -- even if one might argue that because of the extreme smallness of the corrections, it is unnecessary to go to higher orders at all.

The results for massive fields are exponentially suppressed compared to the case of massless fields (as one might have assumed), with the exception of the second Bardeen potential $\Phi_\text{H}$ for a certain range of the non-minimal coupling parameter $\xi$, which shows an enhancement over the massless case. However, this is due to the fact that the correction in the massless case is extremely small for this range of $\xi$, and even vanishes for $\xi = (1 \pm \sqrt{2/5})/6$. The full quantum correction $\Phi_\text{H}^\text{qu}$ is always small, no matter the value of $\xi$. For massless fields, our results can be written in the form of an effective metric for the spinning point particle
\begin{equation}
\label{discussion_metric}
\total s^2 = g_{tt} \total t^2 + g_{rr} \left( \total r^2 + r^2 \total \theta^2 + r^2 \sin^2 \theta \total \phi^2 \right) + 2 g_{t\phi} \total t \total \phi \eqend{,}
\end{equation}
where (reinstating $\hbar$)
\begin{subequations}
\begin{align}
g_{tt} &= - 1 + \frac{2 G_\text{N} M}{r} \left[ 1 + \left[ N_0 \left( 1 + \frac{5}{4} (1-6\xi)^2 \right) + 6 N_{1/2} + 12 N_1 \right] \frac{\hbar G_\text{N}}{45 \pi r^2} \right] \eqend{,} \\
g_{rr} &= 1 + \frac{2 G_\text{N} M}{r} \left[ 1 + \left[ N_0 \left( 1 - \frac{5}{2} (1-6\xi)^2 \right) + 6 N_{1/2} + 12 N_1 \right] \frac{\hbar G_\text{N}}{90 \pi r^2} \right] \eqend{,} \\
g_{t\phi} &= - \frac{2 G_\text{N} M a}{r} \sin^2 \theta \left[ 1 + \left( N_0 + 6 N_{1/2} + 12 N_1 \right) \frac{\hbar G_\text{N}}{20 \pi r^2} \right] \eqend{.}
\end{align}
\end{subequations}
The rotation parameter $a$ is related to the spin $\abs{\vec{S}}$ of the particle by~\eqref{spinning_efe_param_a}
\begin{equation}
a = \frac{\abs{\vec{S}}}{G_\text{N} M} \eqend{,}
\end{equation}
and $N_s$ is the number of massless spin-$s$ fields, with the curvature coupling for scalar fields given by the parameter $\xi$. This could be interpreted as a quantum-corrected linearised Kerr metric, but note that one should not confuse this result with quantum corrections to the exact Kerr metric: first, in our calculation (just as the one of Ref.~\cite{bjerrumbohrdonoghueholstein2003a}) the spinning particle is treated as a test particle in flat spacetime, and the dynamics of quantum fields in spacetimes with horizons, such as the Kerr spacetime, is very different from the flat-space dynamics. Second, even in classical general relativity distributional sources are not acceptable in general~\cite{gerochtraschen1987}, in the sense that the metric that is obtained by solving Einstein's equations with a smeared source and taking the limit where the source becomes point- or line-like depends on the way the limit is taken, if it exists at all. Only in situations where one assumes special symmetry from the outset is such a limit unique and determines a metric fulfilling Einstein's equations with a distributional stress tensor, as has been calculated explicitly for the Schwarzschild, Reissner-Nordstr{\"o}m and Kerr(-Newman) metrics~\cite{balasinnachbagauer1994,kawaisakana1997,balasin1997,pantojarago1997}. Thus, while one could obtain higher-order corrections to our result by taking into account graviton loops, or terms which are of quadratic or higher order in the mass $M$ or the rotation parameter $a$ of the spinning particle (and which are needed in any case to have the proper expansion in $M$ and $a$ of the classical Kerr metric), it is not guaranteed that the result will converge at all, or have the right classical Kerr metric limit.

It seems thus more prudent to stick to a literal interpretation of the calculation, namely quantum corrections to the particle's own gravitational potentials. Note that the particle does not need to be pointlike in reality, but can be an approximation of an extended body, keeping only the first two multipole moments -- mass and spin. In fact, one expects that higher multipole moments, in particular the quadrupole moment, also source a tensor-type potential, which gives quantum corrections to (classical) gravitational radiation. One could then see how these corrections affect the motion of other particles by studying geodesics in the metric~\eqref{discussion_metric}, which, e.g., will give quantum corrections to the motion of heavenly bodies. By studying the motion of particles with spin, it would also be possible to compare with the scattering-type calculations of Ref.~\cite{holsteinross2008}, and the classical results of Ref.~\cite{lalakpokorskiwess1995}. Finally, these calculations should be repeated for other backgrounds, most notably de~Sitter and general Friedmann-Lema{\^\i}tre-Robertson-Walker backgrounds which are relevant for the inflationary period of the early universe. For non-spinning particles and certain types of matter fields, results are already available~\cite{iliopoulosetal1998,wangwoodard2015,parkprokopecwoodard2016,froebverdaguer2016}, and present highly interesting new features, such as quantum corrections which grow logarithmically with either time or distance from the particle, and can thus overcome the small factor $\hbar G_\text{N}$ which suppresses quantum corrections with respect to the classical result.

\begin{acknowledgments}
It is a pleasure to thank Enric Verdaguer for discussions, Ivan Latella for his \texttt{gnuplot} scripts used to create the figures, Daniel Burns and Apostolos Pilaftsis for discussions regarding their work, and Alexei A. Deriglazov for bringing Refs.~\cite{deriglazovramirez2015a,deriglazovramirez2015b} to my attention as well as for discussions on non-minimal spin-gravity coupling. This work is part of a project that has received funding from the European Union’s Horizon 2020 research and innovation programme under the Marie Sk{\l}odowska-Curie grant agreement No. 702750 ``QLO-QG''.
\end{acknowledgments}

\appendix

\section{Metric expansions}
\label{appendix_metric}

Writing a general metric $\tilde{g}_{\mu\nu}$ as background $g_{\mu\nu}$ plus perturbation $h_{\mu\nu}$, we obtain to first order in the perturbation
\begin{subequations}
\begin{align}
\tilde{g}_{\mu\nu} &= g_{\mu\nu} + h_{\mu\nu} \eqend{,} \\
\tilde{g}^{\mu\nu} &= g^{\mu\nu} - h^{\mu\nu} \eqend{,} \\
\sqrt{-\tilde{g}} &= \sqrt{-g} \left( 1 + \frac{1}{2} h \right) \eqend{,} \\
\tilde{\Gamma}^\alpha_{\beta\gamma} &= \Gamma^\alpha_{\beta\gamma} + \frac{1}{2} \left( \nabla_\beta h^\alpha_\gamma + \nabla_\gamma h^\alpha_\beta - \nabla^\alpha h_{\beta\gamma} \right) \eqend{,} \\
\begin{split}
\tilde{R}_{\alpha\beta\gamma\delta} &= R_{\alpha\beta\gamma\delta} + \frac{1}{2} \left( \nabla_\gamma \nabla_{[\beta} h_{\alpha]\delta} - \nabla_\delta \nabla_{[\beta} h_{\alpha]\gamma} + \nabla_\alpha \nabla_{[\delta} h_{\gamma]\beta} - \nabla_\beta \nabla_{[\delta} h_{\gamma]\alpha} \right) \\
&\hspace{4em}- \frac{1}{2} \left( R_{\alpha\beta\mu[\gamma} h_{\delta]}^\mu + R_{\gamma\delta\mu[\alpha} h_{\beta]}^\mu \right) \eqend{,}
\end{split} \\
\tilde{R}_{\alpha\beta} &= R_{\alpha\beta} + \nabla^\delta \nabla_{(\alpha} h_{\beta)\delta} - \frac{1}{2} \nabla^2 h_{\alpha\beta} - \frac{1}{2} \nabla_\alpha \nabla_\beta h \eqend{,} \\
\tilde{R} &= R - h^{\alpha\beta} R_{\alpha\beta} + \nabla^\alpha \nabla^\beta h_{\alpha\beta} - \nabla^2 h \eqend{.}
\end{align}
\end{subequations}
Using the definition of the $n$-dimensional Weyl tensor~\eqref{calculation_weyl_def}, we also obtain
\begin{equation}
\begin{split}
\tilde{C}_{\alpha\beta\gamma\delta} &= C_{\alpha\beta\gamma\delta} + \frac{1}{2} \left( \nabla_\gamma \nabla_{[\beta} h_{\alpha]\delta} - \nabla_\delta \nabla_{[\beta} h_{\alpha]\gamma} + \nabla_\alpha \nabla_{[\delta} h_{\gamma]\beta} - \nabla_\beta \nabla_{[\delta} h_{\gamma]\alpha} \right) \\
&\quad- \frac{1}{n-2} \left( \nabla^\mu \nabla_\alpha h_{\mu[\gamma} + \nabla^\mu \nabla_{[\gamma} h_{\alpha\mu} - \nabla^2 h_{\alpha[\gamma} - \nabla_\alpha \nabla_{[\gamma} h \right) g_{\delta]\beta} \\
&\quad+ \frac{1}{n-2} \left( \nabla^\mu \nabla_\beta h_{\mu[\gamma} + \nabla^\mu \nabla_{[\gamma} h_{\beta\mu} - \nabla^2 h_{\beta[\gamma} - \nabla_\beta \nabla_{[\gamma} h \right) g_{\delta]\alpha} \\
&\quad+ \frac{2}{(n-1)(n-2)} \left( \nabla^\mu \nabla^\nu h_{\mu\nu} - \nabla^2 h \right) g_{\alpha[\gamma} g_{\delta]\beta} - \frac{1}{2} \left( R_{\alpha\beta\mu[\gamma} h_{\delta]}^\mu + R_{\gamma\delta\mu[\alpha} h_{\beta]}^\mu \right) \\
&\quad- \frac{2}{(n-1)(n-2)} \left[ (n-1) R_{\alpha[\gamma} - R g_{\alpha[\gamma} \right] h_{\delta]\beta} \\
&\quad+ \frac{2}{(n-1)(n-2)} \left[ (n-1) R_{\beta[\gamma} - R g_{\beta[\gamma} \right] h_{\delta]\alpha} - \frac{2}{(n-1)(n-2)} h^{\mu\nu} R_{\mu\nu} g_{\alpha[\gamma} g_{\delta]\beta} \eqend{.}
\end{split}
\end{equation}

\section{The master integral}
\label{appendix_master}

We want to calculate the integral
\begin{equation}
I_z(\vec{x}) \equiv \int \left[ (x_{++}^2)^{z-1} - (x_{+-}^2)^{z-1} \right] \total t
\end{equation}
for $-1 < \Re z < 0$, where the different prescriptions are defined by equation~\eqref{calculation_x2_prescriptions}. As explained after equation~\eqref{calculation_stress_tensor_combi}, the integrand vanishes unless $(t,\vec{x})$ is in the backward lightcone emanating from the origin $(0,\vec{0})$. Especially, it vanishes for $t > 0$, and inserting the explicit form of the prescriptions~\eqref{calculation_x2_prescriptions} we thus obtain
\begin{equation}
I_z(\vec{x}) = \lim_{\epsilon \to 0} \int_{-\infty}^0 \left[ \left[ r^2 - ( t + \mathi \epsilon )^2 \right]^{z-1} - \left[ r^2 - ( t - \mathi \epsilon )^2 \right]^{z-1} \right] \total t \eqend{.}
\end{equation}
with $r \equiv \abs{\vec{x}}$. An indefinite integral is given by
\begin{equation}
\int \left[ r^2 - ( t \pm \mathi \epsilon )^2 \right]^{z-1} \total t = r^{2z-2} ( t \pm \mathi \epsilon ) \hypergeom{2}{1}\left( \frac{1}{2}, 1-z: \frac{3}{2}; \frac{( t \pm \mathi \epsilon )^2}{r^2} \right)
\end{equation}
with the Gau{\ss} hypergeometric function $\hypergeom{2}{1}$, as can be checked directly from its series definition. By a standard hypergeometric transformation~\cite{dlmf}, we bring it into the form
\begin{equation}
\begin{split}
&r^{2z-2} ( t \pm \mathi \epsilon ) \frac{\sqrt{\pi} \Gamma\left( \frac{1}{2} - z \right)}{2 \Gamma(1-z)} \left( - \frac{( t \pm \mathi \epsilon )^2}{r^2} \right)^{-\frac{1}{2}} \\
&\quad+ r^{2z-2} ( t \pm \mathi \epsilon ) \frac{1}{2z-1} \left( - \frac{( t \pm \mathi \epsilon )^2}{r^2} \right)^{-1+z} \hypergeom{2}{1}\left( 1-z, \frac{1}{2}-z; \frac{3}{2}-z; \frac{r^2}{( t \pm \mathi \epsilon )^2} \right) \eqend{,}
\end{split}
\end{equation}
which is suitable for taking the lower limit $t \to - \infty$. Since $\Re z < 0$, the second term does not contribute in this limit, and carefully evaluating the inverse square root in the first term for the different prescriptions we obtain
\begin{equation}
\label{appendix_master_result_z}
I_z(\vec{x}) = \lim_{\epsilon \to 0} \left[ 2 r^{2z-2} \mathi \epsilon \hypergeom{2}{1}\left( \frac{1}{2}, 1-z: \frac{3}{2}; - \frac{\epsilon}{r^2} \right) - \mathi r^{2z-1} \frac{\sqrt{\pi} \Gamma\left( \frac{1}{2} - z \right)}{\Gamma(1-z)} \right] = - \mathi r^{2z-1} \frac{\sqrt{\pi} \Gamma\left( \frac{1}{2} - z \right)}{\Gamma(1-z)} \eqend{.}
\end{equation}

For the massless case, we also need the integral with $\ln(\mu^2 x^2)/x^2$, which can be obtained as
\begin{equation}
\frac{\ln(\mu^2 x^2)}{x^2} = \lim_{\delta \to 0} \frac{1}{\delta} \left[ \mu^{-2\delta} (x^2)^{-1-\delta} - \mu^{-4\delta} (x^2)^{-1-2\delta} \right]
\end{equation}
in such a way to ensure $\Re z < 0$. Thus it follows that
\begin{equation}
\label{appendix_master_result_log}
\int \left[ \frac{\ln(x_{++}^2)}{x_{++}^2} - \frac{\ln(x_{+-}^2)}{x_{+-}^2} \right] \total t = \lim_{\delta \to 0} \frac{\mu^{-2\delta} I_{-\delta}(\vec{x}) - \mu^{-4\delta} I_{-2\delta}(\vec{x})}{\delta} = - 2 \pi \mathi \frac{\ln(2 \mu r)}{r} \eqend{.}
\end{equation}

\section{Asymptotic expansion}
\label{appendix_asymptotic}

We want to obtain an asymptotic expansion as $m r \to \infty$ of an integral of the form
\begin{equation}
\label{appendix_asymptotic_int}
I_a(m r) \equiv \int_\mathcal{C} (m r)^{2z} \frac{\Gamma(-z) \Gamma(a-z) \Gamma\left( \frac{3}{2} - z \right)}{\Gamma\left( \frac{7}{2} - z \right)} \frac{\total z}{2\pi\mathi} \eqend{,}
\end{equation}
where $a \geq 0$, and the contour $\mathcal{C}$ runs from $- \mathi \infty$ to $+ \mathi \infty$ with $\Re z < 0$. If the integrand would contain $\Gamma$ functions with poles in the left half-plane, of the form $\Gamma(b+z)$, we could shift the contour over the poles at $z = -b-k$, and obtain an asymptotic expansion of the integral in the form of the corresponding residues $\sim (m^2 r^2)^{-b-k}$. However, in our case we can shift the contour to arbitrary $\Re z < 0$, and thus $I_a(m r)$ decays faster than any polynomial in $m r$ as $m r \to \infty$, which is a signal of an exponentially small asymptotic expansion. As explained below, the order of this expansion is essentially controlled by the multiplicity and the shift of the $\Gamma$ functions appearing in the integrand. In our case, this is equal to $2$ and $a-5/2$, respectively, and we thus expect the leading term of the expansion to be given by $\mathe^{- 2 m r} (2mr)^{a - 5/2}$.

To obtain the corresponding asymptotic expansion, we want to bring the integrand into a form where we can use the integral
\begin{equation}
\label{appendix_asymptotic_exp}
\int_\mathcal{C} u^{-z} \Gamma(z-b) \frac{\total z}{2\pi\mathi} = u^{-b} \mathe^{-u} \eqend{,}
\end{equation}
where the contour $\mathcal{C}$ runs from $- \mathi \infty$ to $+ \mathi \infty$ with $\Re z > \Re b$. This can be done using so-called inverse factorial expansions~\cite{olver}, for which we need the well-known asymptotic expansion of the $\Gamma$ function
\begin{equation}
\ln \Gamma(z) = \left( z - \frac{1}{2} \right) \ln z - z + \frac{1}{2} \ln(2\pi) + \sum_{k=1}^n \frac{B_{2k}}{2k (2k-1) z^{2k-1}} + \bigo{\abs{z}^{-2n-1}}
\end{equation}
with the Bernoulli numbers $B_{2k}$. Since this expansion is not valid near the negative real axis, we first have to change our integration variable $z \to -z$, obtaining
\begin{equation}
\label{appendix_asymptotic_iposz}
I_a(m r) = \int_\mathcal{C} (m r)^{-2z} \frac{\Gamma(z) \Gamma(a+z) \Gamma\left( \frac{3}{2} + z \right)}{\Gamma\left( \frac{7}{2} + z \right)} \frac{\total z}{2\pi\mathi} \eqend{.}
\end{equation}
The contour $\mathcal{C}$ now has $\Re z > 0$, and since there are no poles in the right half-plane we can shift the contour to have $\Re z \gg 1$. We then have to choose parameters $\mu$ (the multiplicity) and $\nu$ (the shift) such that the sum of the expansions of $- \ln \Gamma(\mu z + \nu)$ and the $\Gamma$ functions in the integrand does not contain any term $\sim \ln z$. In the case at hand, these are given by $\mu = 2$ and $\nu = a - 5/2$, and we content ourselves with an expansion up to next-to-leading order. Therefore, we get
\begin{equation}
\begin{split}
&\ln \Gamma(z) + \ln \Gamma(a+z) + \ln \Gamma\left( \frac{3}{2} + z \right) - \ln \Gamma\left( \frac{7}{2} + z \right) - \ln \Gamma\left( 2 z + a - \frac{5}{2} \right) \\
&\quad= - 2 z \ln 2 + (3-a) \ln 2 + \frac{1}{2} \ln(2\pi) + \frac{4 a^2 + 16 a - 97}{16z} + \bigo{z^{-2}} \eqend{,}
\end{split}
\end{equation}
and exponentiating it follows that
\begin{equation}
\label{appendix_asymptotic_gamma}
\begin{split}
\frac{\Gamma(z) \Gamma(a+z) \Gamma\left( \frac{3}{2} + z \right)}{\Gamma\left( \frac{7}{2} + z \right)} &= \Gamma\left( 2 z + a - \frac{5}{2} \right) \sqrt{2\pi} \, 2^{3-a-2z} \left[ 1 + \frac{4 a^2 + 16 a - 97}{16 z} + \bigo{z^{-2}} \right] \\
&= \sqrt{2\pi} \, 2^{3-a-2z} \Bigg[ \Gamma\left( 2 z + a - \frac{5}{2} \right) + \frac{4 a^2 + 16 a - 97}{8} \Gamma\left( 2 z + a - \frac{7}{2} \right) \\
&\hspace{10em}+ \bigo{1} \Gamma\left( 2 z + a - \frac{9}{2} \right) \Bigg] \eqend{.} \raisetag{1.5\baselineskip}
\end{split}
\end{equation}
We can now insert this expansion into the integral~\eqref{appendix_asymptotic_iposz} and use equation~\eqref{appendix_asymptotic_exp} to integrate each term. Since the multiplicity $\mu \neq 1$, we have to rescale the integration variable $z$ first, and this together with the explicit factor of $2^{-2z}$ in equation~\eqref{appendix_asymptotic_gamma} gives the exponential falloff $\sim \mathe^{-2 m r}$. Moreover, equation~\eqref{appendix_asymptotic_exp} shows that the leading power of $m r$ is directly given by the shift $\nu$.

Taking everything together, it follows that
\begin{equation}
\label{appendix_asymptotic_expansion}
I_a(m r) = \sqrt{\pi} \, \mathe^{-2 m r} (m r)^{a - \frac{5}{2}} \left[ 1 + \frac{4 a^2 + 16 a - 97}{16 m r} + \bigo{ \frac{1}{m^2 r^2}} \right] \eqend{,}
\end{equation}
which is the desired asymptotic expansion.

\addcontentsline{toc}{section}{References}
\bibliography{literature}

\end{document}